\documentclass[usenatbib]{mn2e}
\usepackage{graphicx}

\def\apj{\rm ApJ}
\def\apjl{\rm ApJL}
\def\apjs{\rm ApJS}
\def\aj{\rm AJ}
\def\mnras{\rm MNRAS}
\def\nat{\rm Nature}

\def\pasp{\rm PASP}
\def\pasa{\rm PASA}
\def\aap{\rm AAP}
\def\araa{\rm ARA\&A}

\def\prd{\rm PRD}

\def\physrep{\rm Physics Reports}

\def\gax{\mathrel{\raise.3ex\hbox{$>$}\mkern-14mu\lower0.6ex\hbox{$\sim$}}}
\def\lax{\mathrel{\raise.3ex\hbox{$<$}\mkern-14mu\lower0.6ex\hbox{$\sim$}}}
\def\gtorder{\mathrel{\raise.3ex\hbox{$>$}\mkern-14mu
             \lower0.6ex\hbox{$\sim$}}}
\def\ltorder{\mathrel{\raise.3ex\hbox{$<$}\mkern-14mu
             \lower0.6ex\hbox{$\sim$}}}

\begin{document}

\title [Supernova Progenitors and Their Variability]
   {Supernova Progenitors, Their Variability, and the Type~IIP Supernova ASASSN-16fq in M~66}

\author[C.~S. Kochanek et al.]{ 
    C.~S. Kochanek$^{1,2}$, 
    M. Fraser$^{3}$, 
    S.~M. Adams$^{4}$, 
    T.~Sukhbold$^{1,2}$, 
    J.~L. Prieto$^{5,6}$, \newauthor 
    T. M\"uller$^{6,7}$,
    G. Bock$^{8}$,
    J.~S. Brown$^{1}$, 
    Subo Dong$^{9}$,
    T.~W.-S. Holoien$^{1}$,
    R. Khan$^{10}$, \newauthor 
    B.~J. Shappee$^{11,12}$, 
    K.~Z. Stanek$^{1,2}$
    \\
  $^{1}$ Department of Astronomy, The Ohio State University, 140 West 18th Avenue, Columbus OH 43210 \\
  $^{2}$ Center for Cosmology and AstroParticle Physics, The Ohio State University,
    191 W. Woodruff Avenue, Columbus OH 43210 \\
  $^{3}$ Institute of Astronomy, University of Cambridge, Madingley Rd, Cambridge CB3 0HA, UK \\
  $^{4}$ Cahill Center for Astrophysics, California Institute of Technology, Pasadena, CA 91125, USA \\
  $^{5}$ N\'ucleo de Astronom\'ia de la Facultad de Ingenier\'ia, Universidad Diego Portales, Av. Ej\'ercito 441, Santiago, Chile \\
  $^{6}$ Millennium Institute of Astrophysics, Santiago, Chile \\
  $^{7}$ Instituto de Astrof\'isica, Pontifica Universidad Cat\'olica de Chile, Av. Vicu\~na Mackenna 4860, 782-0436 Macul, Santiago, Chile \\
  $^{8}$ Runaway Bay Observatory, Australia \\
  $^{9}$ Kavli Institute for Astronomy and Astrophysics, Peking University, Yi He Yuan Road 5, Hai Dian District, Beijing 100871, China \\
  $^{10}$ Department of Astronomy, Box 351580, University of Washington, Seattle, WA 98195 \\
  $^{11}$ Carnegie Observatories, 813 Santa Barbara Street, Pasadena, CA 91101, USA \\
  $^{12}$ Hubble, Carnegie-Princeton Fellow \\
   }

\maketitle

\begin{abstract}
We identify a pre-explosion counterpart to the nearby Type~IIP supernova ASASSN-16fq (SN~2016cok) 
in archival Hubble Space Telescope (HST) data.  The source
appears to be a blend of several stars that prevents obtaining accurate photometry. However,
with reasonable assumptions about the stellar temperature and extinction, the progenitor
almost certainly had an initial mass $M_* \ltorder 17 M_\odot$ and was most likely in the
mass range $M_*=8$-$12M_\odot$.  Observations once ASASSN-16fq has faded 
will have no difficulty accurately determining the properties of the progenitor.  
In 8 years of Large Binocular Telescope (LBT) data, 
no significant progenitor variability is detected to RMS limits of roughly 0.03~mag.
Of the six nearby SN with constraints on low level variability, SN~1987A,
SN~1993J, SN~2008cn, SN~2011dh, SN~2013ej and ASASSN-16fq, only the slowly fading progenitor
of SN~2011dh showed clear evidence of variability.  Excluding SN~1987A, the 90\% confidence
limit implied by these sources on the number of outbursts over the last decade before
the SN that last longer than 0.1~years (FWHM) and are brighter than $M_R<-8$ mag 
is approximately $N_{out} \ltorder 3$.  Our continuing LBT monitoring program 
will steadily improve constraints on pre-SN progenitor variability at
amplitudes far lower than achievable by SN surveys.
\end{abstract}


\section{Introduction}
\label{sec:introduction}

At the end of their lives, all massive ($\gtorder 8M_\odot$) stars
must undergo core collapse once their iron cores become too 
massive to be stable.  In most cases, this leads to a supernova
(SN) explosion probably driven by some combination of neutrino heating
and the effects of turbulence and convection (see the recent review
by \cite{Muller2016} and, e.g., recent
results by \citealt{Couch2015}, \citealt{Dolence2015}, \citealt{Wongwathanarat2015}).  The visible
properties of the successful SNe then depend on the degree
of mass loss, ranging from Type~IIP SN which have retained
most of their hydrogen envelopes, to Type~Ic SN which appear
to have been stripped even of helium (e.g., \citealt{Filippenko1997}).  The mass loss is 
controlled by some combination of intrinsic effects such
as winds and extrinsic effects such as binary mass transfer (see the
review by \citealt{Smith2014b}).

There is no strong requirement that more than roughly 50\% of core
collapses lead to successful SN (e.g., neutrino backgrounds, \citealt{Lien2010};
star formation rates, \citealt{Horiuchi2011}; nucleosynthesis, \citealt{Brown2013},
\citealt{Clausen2015}) 
and a 10-30\% fraction of failed SN producing black holes 
without a dramatic external
explosion is both expected in many modern analyses of the
``explodability'' of stars (e.g., \citealt{Ugliano2012},
\citealt{OConnor2013},
\citealt{Pejcha2015}, \citealt{Ertl2016}, \citealt{Sukhbold2016})
and would provide a natural
explanation of the compact remnant mass function (\citealt{Kochanek2014},
\citealt{Kochanek2015}, \citealt{Clausen2015}).  Indeed,
scenarios for the recent gravitational wave detection of
a merging black hole binary (\citealt{Abbott2016}) 
all invoke at least one failed SN (e.g., \citealt{Belczynski2016},
\citealt{Woosley2016}).

A powerful means of probing these issues is to work out the
mapping between successful SNe and their progenitor stars.
This is a challenging observational program (see the reviews
by \citealt{Smartt2009} and  \citealt{Smartt2015})
which has slowly been carried out over the last 
20 years (e.g., \citealt{VanDyk2003},
\citealt{Smartt2004}, \citealt{Li2006},
\citealt{Smartt2009b}, \citealt{EliasRosa2009}, \citealt{EliasRosa2011},
\citealt{Maund2011}, \citealt{VanDyk2011}, \citealt{Fraser2012},
\citealt{Fraser2014}).
With one possible exception (\citealt{Cao2013}, \citealt{Folatelli2016}, see \cite{Eldridge2013}
for a discussion of limits), all the 
identified progenitors are of Type~II (IIP, IIL, IIb, or IIn).

As first pointed out by \cite{Kochanek2008} and then
better quantified by \cite{Smartt2009b}, there appears
to be a deficit of higher mass SN progenitors.  In particular,
\cite{Smartt2009b} only identified Type~IIP progenitors with
masses of $\ltorder 17M_\odot$ even though stars up to $25$-$30M_\odot$
are expected to explode as red supergiants with most of their
hydrogen envelopes.  While attempts have been made to explain
this using extinction by winds (\citealt{Walmswell2012}, but
see \citealt{Kochanek2012}) or by modifying stellar 
evolution (e.g. \citealt{Groh2013}), the same problem of missing, higher mass progenitors
is seen in examinations of the stellar populations near local
group SN remnants (\citealt{Jennings2014}).  \cite{Jerkstrand2014}
also argue that no Type~IIP SN have shown nucleosynthetic evidence
for a higher mass ($M_*>20M_\odot$) progenitor. Following the proposal
of \cite{Kochanek2008}, \cite{Gerke2015}
have been carrying out a search for failed SN with the
Large Binocular Telescope (LBT), identifying one 
promising candidate (see also \citealt{Reynolds2015}).  The progenitor of this candidate for
a failed SN appears to be a red supergiant in exactly the
mass range missing from searches for the progenitors of
successful SN (Adams et al. 2016, in preparation).  

\begin{figure*}
\centering
\includegraphics[width=\textwidth]{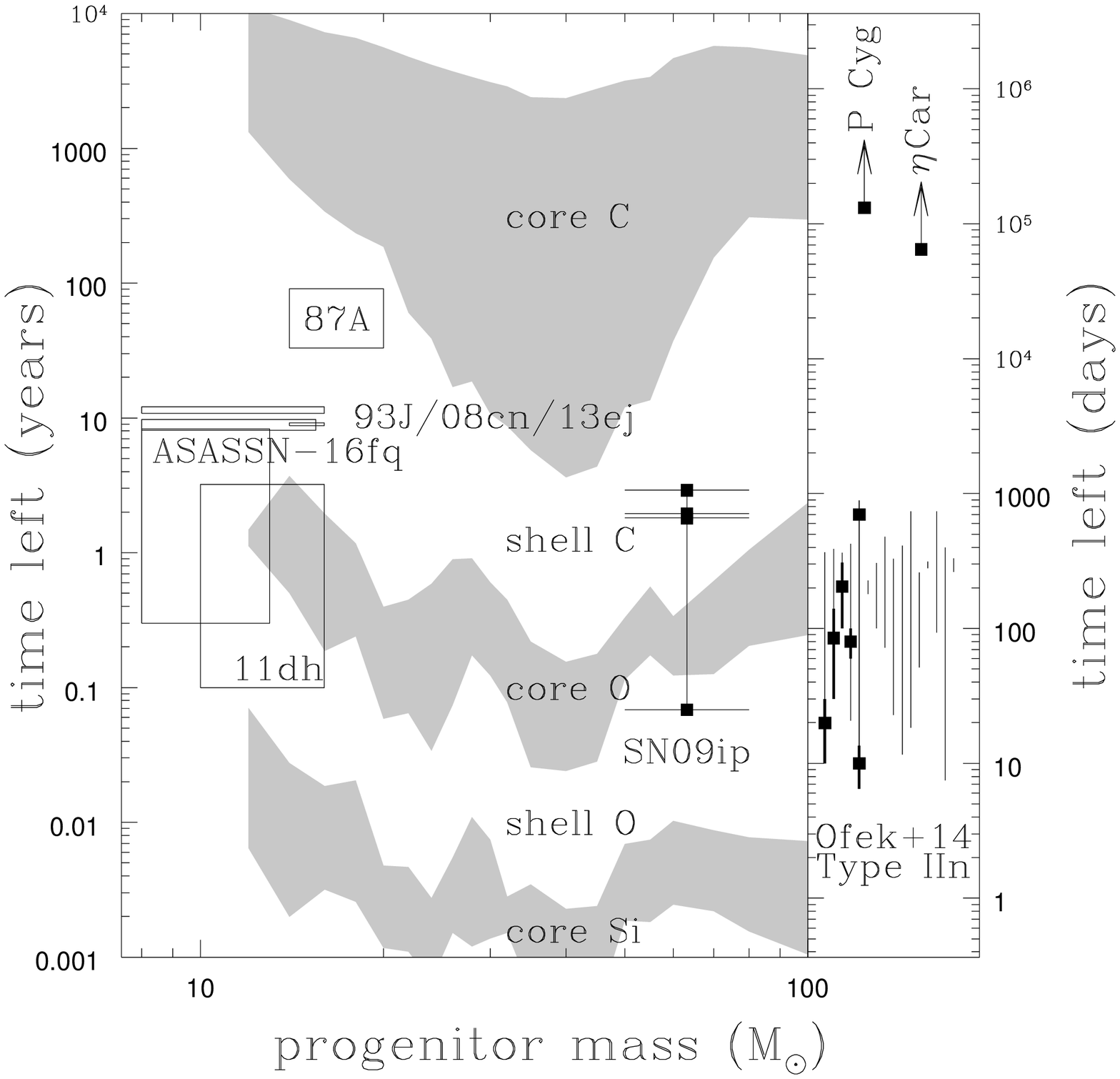}
\caption{Final nuclear burning stages as a function of progenitor mass based on the
standard, non-rotating, $12$-$100M_\odot$, solar metallicity models of \protect\cite{Sukhbold2014} and
\protect\cite{Woosley2007}.  The gray bands show, from
top to bottom, the periods of core carbon, oxygen and silicon burning, separated by periods of 
shell burning.  
The points associated with SN~2009ip indicate the timing of its outbursts relative to its
presumed explosion along with its estimated mass range.  The sub-panel to the right shows
16 thin vertical lines for the (control) time periods sampled by PTF for 16 Type~IIn SN, 
with heavy black points and lines for the time periods associated with outbursts 
(\protect\citealt{Ofek2014}).  The masses of the progenitors of these stars are unknown
but they are generally assumed to be large.  For comparison, the sub-panel also
indicates the present day lower limits for the 1840 and 1655 outbursts of $\eta$~Car
and P~Cyg.   The boxes at lower masses show the progenitor mass ranges
and the time periods that can be surveyed for progenitor variability for SN~1987A, SN~1993J,
SN~2008cn, SN~2011dh, SN~2013ej, and ASASSN-16fq.  For a Salpeter IMF with SN occurring in 
the mass range from $8$ to $100M_\odot$, 50\% of SN arise from the mass range from $8$ to
$13.1M_\odot$.
  }
\label{fig:bigpicture}
\end{figure*}

A second recent puzzle about SN progenitors is that some
appear to have outbursts (\citealt{Pastorello2007}, \citealt{Fraser2013},
\citealt{Mauerhan2013},
\citealt{Ofek2014}, \citealt{Ofek2016}) and/or eject significant amounts
of mass (see \citealt{Galyam2012}, \citealt{Smith2014b}) shortly before 
they explode.  The most extreme mass-loss events ($\dot{M} \sim M_\odot$/year)
likely explain the rare, superluminous Type~IIn SNe (\citealt{Smith2007}),
but the inferred mass loss rates are frequently $\dot{M} \gtorder 10^{-3}M_\odot$/year
even for normal Type IIn SNe (see, e.g., \citealt{Kiewe2012}).  The local
systems known to reach such extreme mass loss rates are the 
Luminous Blue Variables (LBVs), with $\eta$~Carinae as the
most spectacular example (see \citealt{Humphreys1994}).
The rate of $\eta$~Carinae-like events is roughly 10\% of the SN 
rate (\citealt{Kochanek2011}, \citealt{Khan2015a}, \citealt{Khan2015b}), 
which is sufficient to explain the occurrence of the extreme
Type~IIn superluminous SN.  Any association of LBV eruptions
with the very late phases of stellar evolution would roughly 
require the typical $M_* \gtorder 50 M_\odot$ star to have
at least one eruption in the $\sim 10^3$~year period 
after carbon ignition (\citealt{Kochanek2011}). On the other hand,
theoretical models to explain pre-SN outbursts and
Type~IIn SNe have favored mechanisms associated with
the last few years, corresponding to the neon/oxygen burning phases or later
(\citealt{Quataert2012}, \citealt{Shiode2014}, \citealt{Smith2014},
\citealt{Woosley2015}). 
In this picture, massive stars must have two separate mechanisms for 
triggering outbursts, one to explain
the LBVs and a second to explain the pre-SN outbursts.

The existence of any transients associated with
the last $\ltorder 10^3$~years (or less) of stellar life 
requires a causal mechanism associated with these final phases
(see the discussion in \citealt{Kochanek2011}). Figure~\ref{fig:bigpicture}
shows the dependence of the final nuclear burning stages on 
progenitor mass for the standard, non-rotating, $12$-$100M_\odot$, 
solar metallicity models of \cite{Sukhbold2014} and
\protect\cite{Woosley2007}.  We show the periods of core
and shell carbon, oxygen
and silicon burning -- the neon burning phase is not as energetically
important.  The large scale structure
in Figure~\ref{fig:bigpicture}, with the shortest time scales for 
intermediate masses, is driven by the rapid increase in mass loss for
the higher mass stars.  The smaller scale variations in the mass-dependence
of the post-carbon burning phases are due to the complex interplay of the burning phases 
and their consequences for structure of the stellar core
(see \cite{Sukhbold2014} for a detailed discussion).  

We illustrate the outbursts associated with Type~IIn SN
in Figure~\ref{fig:bigpicture} by SN~2009ip and the Palomar 
Transient Factory (PTF) sample of Type~IIn SN considered by 
\cite{Ofek2014}.  SN~2009ip has an estimated progenitor mass
of $50$-$80M_\odot$ (\citealt{Smith2010}) and showed a series
of outbursts before the apparent explosion (see, e.g., \citealt{Smith2010},   
\citealt{Foley2011}, \citealt{Mauerhan2013}, \citealt{Pastorello2013}, 
\citealt{Margutti2014}).  For the PTF sample, the progenitor
masses are unknown.  PTF data are available for the last few years
before the SN, as shown by the lines spanning the survey times
for each SN. \cite{Ofek2014} detect 5 outbursts and argue that 
it is highly probable that all Type~IIn SN experience outbursts
and that many are simply missed due to the survey
depth and cadence.  The outbursts shown in Figure~\ref{fig:bigpicture}
are associated with the very last phases of carbon shell burning 
through the early phases of oxygen shell burning.  It seems 
probable, particularly in the case of SN~2009ip, that outbursts
cannot be restricted to the time period after the initiation of
core oxygen burning.  As a contrast,
if the eruption mechanism of LBVs had any correlation with these
last phases, it would have to be associated with the carbon 
burning phase, as illustrated in Figure~\ref{fig:bigpicture} by
the 1840 and 1655 outbursts of $\eta$ Car and P Cygni (see
\citealt{Humphreys1994}). 

Broadly speaking, there are two possible scenarios associated
with these pre-SN transients.  The first option is that only the high amplitude
events seen in the SN surveys or implied by the 
Type~IIn SNe exist and they are associated with a very narrow range
of progenitor parameter space (e.g. mass, metallicity, rotation).
The second option is that the outburst mechanism is relatively generic,
and the existing events simply represent the high amplitude
tail of a much broader distribution. Unfortunately,
the existing systematic searches for outbursts 
(e.g. \citealt{Ofek2014},
\citealt{Bilinski2015}, \citealt{Strotjohann2015})  
are all part of searches
for supernovae and essentially cannot detect significantly
lower amplitude transients.

Like building the mapping between
SN and progenitors, building the mapping between pre-SN 
outbursts and progenitors requires surveys of much greater
sensitivity than searches for SN.  Unfortunately, where data deep enough to 
observe progenitors are already rare, having multiple epochs
of such data to study progenitor variability is rarer still.
At present, such data only exists for the progenitors
of SN~1987A (see \cite{Plotkin2004} and references
therein), SN~1993J (\citealt{Cohen1995}),
SN~2008cn (\citealt{EliasRosa2009}, \citealt{Maund2015}),
SN~2011dh (\citealt{Szczygiel2012}), and SN~2013ej (\citealt{Fraser2014}). 
These sources all have progenitor detections and mass estimates, placing
them below $20 M_\odot$. Figure~\ref{fig:bigpicture} shows the
region of progenitor mass and remaining life time the data can probe.
The variability constraints
for SN~1987A and SN~1993J are relatively poor
and only SN~2011dh shows clear evidence for low levels of variability.
All these systems are also in the $\ltorder 20M_\odot$ mass range
suggested by \cite{Shiode2014} for wave-driven mass loss at solar
metallicity and some are likely near the $\sim 10M_\odot$ mass range
associated with the explosive silicon burning mechanism of 
\cite{Woosley2015}.

\begin{figure*}
\begin{center}
\includegraphics[width=0.45\textwidth]{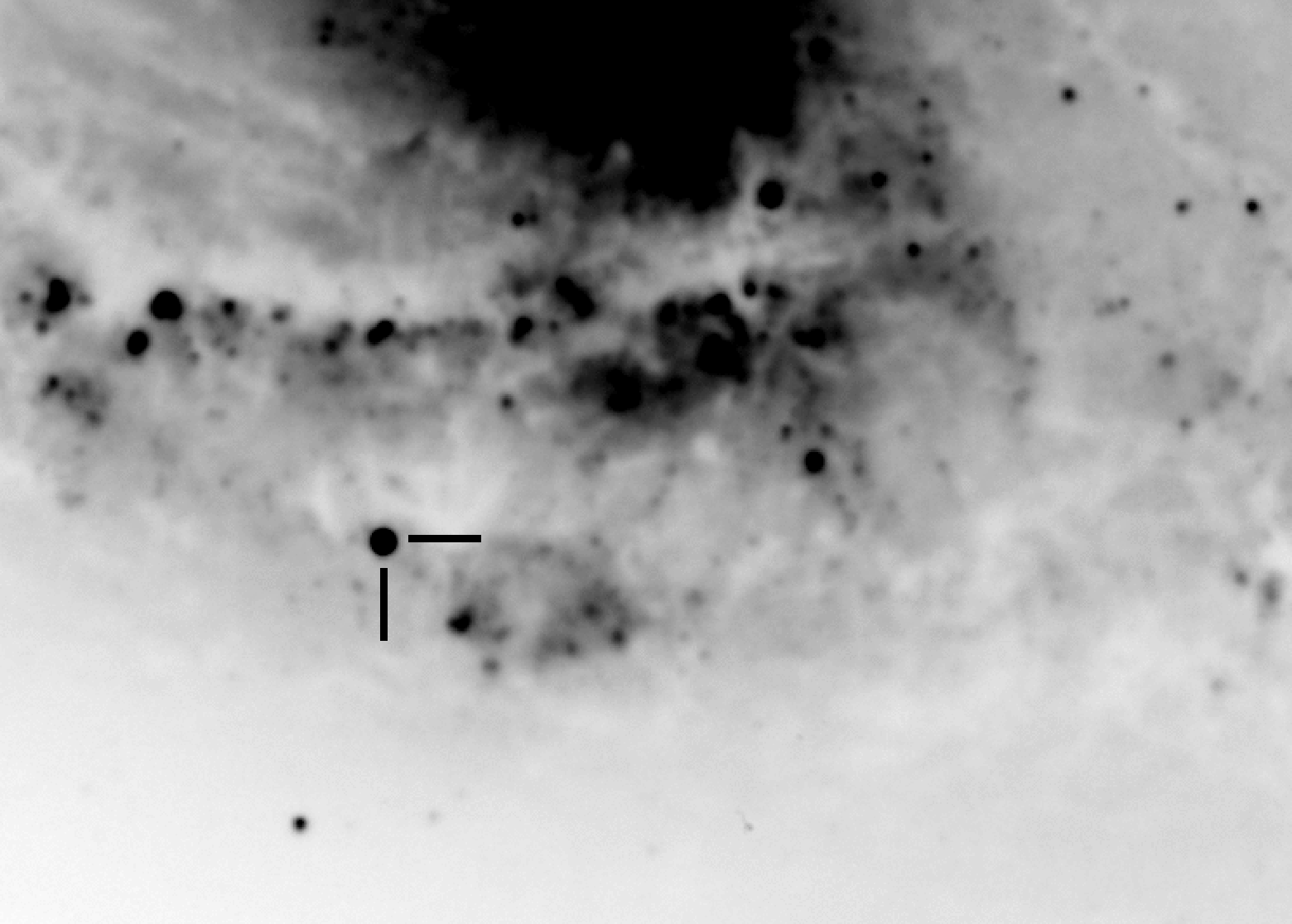}
\includegraphics[width=0.45\textwidth]{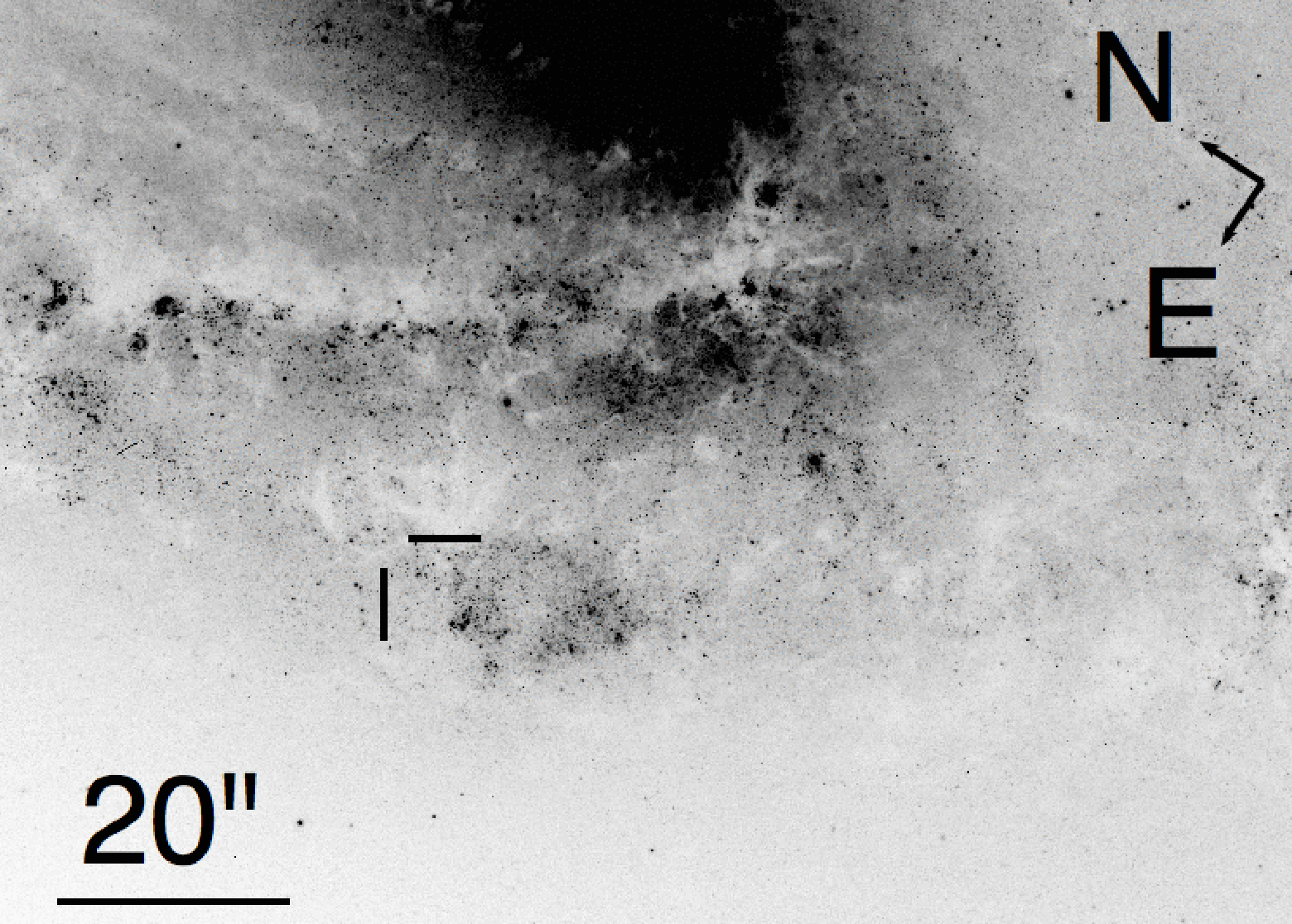}\llap{\raisebox{0.0cm}{\includegraphics[width=2.5cm,height=2.5cm]{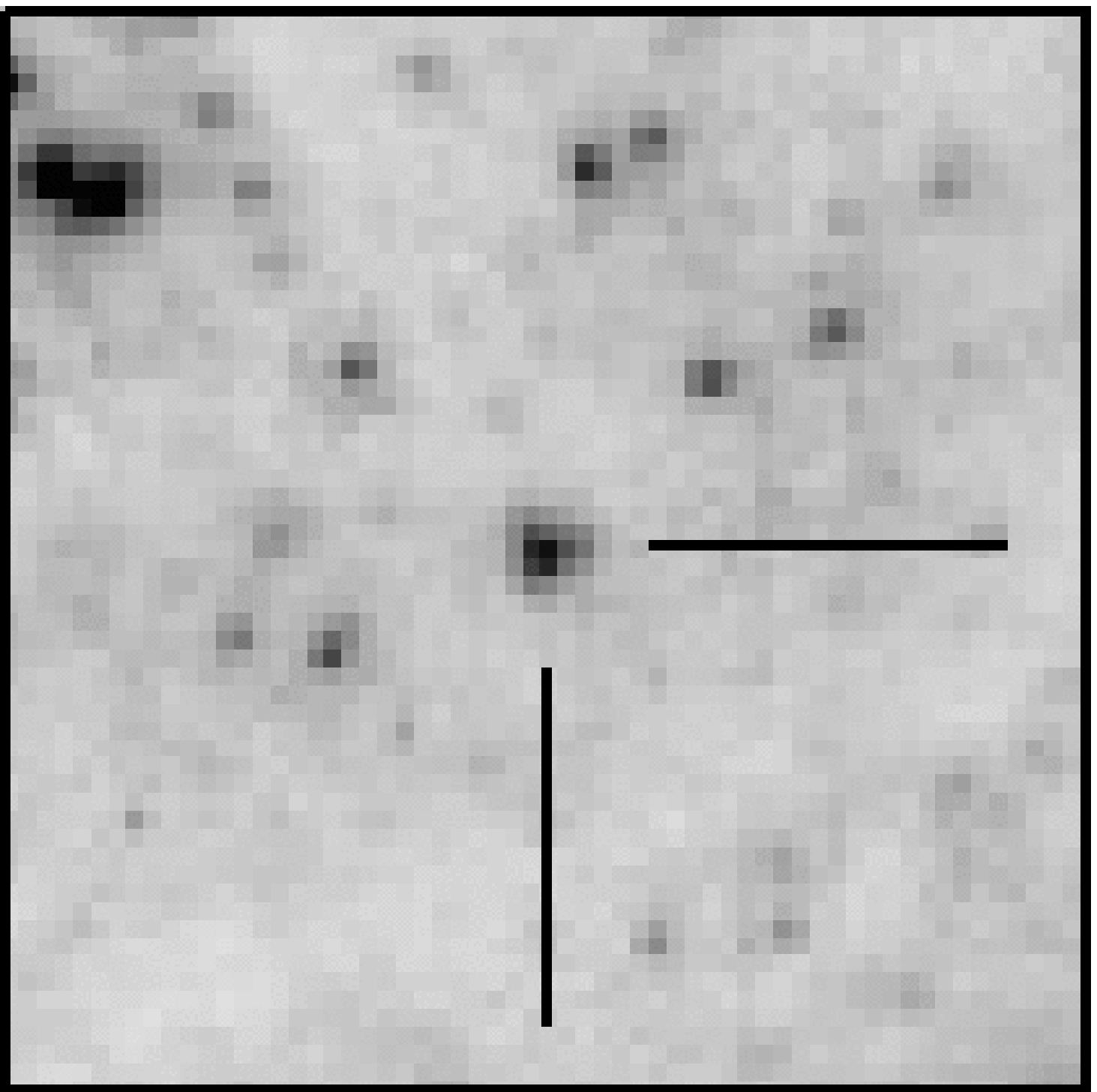}}}
\end{center}
\caption{Identification of a progenitor candidate for ASASSN-16fq. The left panel shows an image of the
SN taken with LBT, aligned to the pre-explosion HST ACS/WFC F814W image shown in the right panel.
The position of the SN is indicated in both panels. The inset in the right, pre-explosion panel
shows a 3\farcs0$\times$3\farcs0 region centered on the progenitor candidate, with the derived
SN position indicated.}
\label{fig:progenitor}
\end{figure*}

Here we report on the properties of the progenitor of 
ASASSN-16fq (SN~2016cok).  ASASSN-16fq was discovered
(\citealt{Bock2016}) in NGC~3627 (M~66) by the All-Sky Automated Survey for 
Supernovae (ASAS-SN, \citealt{Shappee2014}) on 28 May 2016 and was spectroscopically 
classified as a Type~IIP SN (\citealt{Zhang2016}).  There are multiple
epochs of HST data because of the debated transient 
SN~1997bs (\citealt{VanDyk2000}, \citealt{Smith2011},
\citealt{Kochanek2012b}, \citealt{Adams2015}) 
and the Type~IIL SN~2009hd (\citealt{EliasRosa2011},
\citealt{Tinyanont2016}).  It
is also one of the galaxies monitored as part of the
search for failed supernovae with the LBT
(\citealt{Kochanek2008}, \citealt{Gerke2015}), allowing a
deep search for progenitor variability over its last
8 years (see Figure~\ref{fig:bigpicture}). 
In \S2 we identify and describe the
progenitor primarily based on archival HST data to
make a rough estimate of its luminosity
and initial mass.  In \S3, we search for variability
from the progenitor 
using the data from the LBT.  We discuss the results in \S4,
focusing on an extended discussion of supernova progenitor variability.  
Following \cite{Gerke2015}, we adopt a distance of
$10.62$~Mpc from \cite{Kanbur2003} and a Galactic extinction of 
of $E(B-V)=0.03$~mag from \cite{Schlafly2011}.

\section{Identification and Properties of the Progenitor}

The region including ASASSN-16fq was observed by HST 
on 28 December 1994 (WFPC2/F606W, GO-54456, PI Illingworth),
26 November 2001 (WFPC2/F606W, GO-8597, PI Regan),
24 February and 4 March 2001 (WFPC2/F555W and F814W, GO-8602, PI Filippenko),
30/31 December 2004 (ACS/F658N and F435W, GO-10402, PI Chandar),
14 December 2009 (ACS/F555W and F814W, GO-11575, PI Van Dyk),
and 28 November 2013 (WFC3/UVIS/F555W and F814W, GO-13477, PI Kochanek).
These data are summarized in Table~\ref{tab:hstphot}.
The region has also been observed multiple times by Spitzer at
$3.6$ and $4.5\mu$m (programs 159, 10001 and 10136/11063,
PIs Kennicutt, Kochanek and Kasliwal, respectively). 

In order to determine the position of the SN in the HST images,
we obtained new LBT data including the SN consisting 
of 24 five second R-band exposures with a 1\farcs05
full width at half-maximum (FWHM) and a nominal 
R-band depth when combined of roughly $24.7$~mag ($S/N\simeq 5$).  
We identified 39 sources
in common between the combined LBT image and the pipeline,
drizzled, CTE-corrected (charge transfer efficiency) ACS/WFC
F814W HST image taken on 2009 December 14.  A pixel 
coordinate transformation allowing for rotation, translation,
independent $x$ and $y$ pixel scalings and   
second order $x^2$, $xy$ and $y^2$ terms to account for 
distortions ($24$ coefficients
in total) led to a geometric transformation with root-mean-square
(RMS) errors in the $x$ and $y$ HST pixel axes of $0.164$ and
$0.169$ LBT pixels, respectively.  The position of the SN in the
LBT image was measured using three different centering 
algorithms which agreed to $\ltorder 0.02$ LBT pixels
(pixel scale $0\farcs226$).  The resulting estimate of the pixel position
of the SN on the F814W image is $(1658.047\pm0.734, 2014.933\pm0.757)$. 
Figure~\ref{fig:progenitor} shows the
LBT image with the SN and the same region in the pre-explosion
HST image along with an inset showing a 3\farcs0-square region
centered on the estimated position of the progenitor.  A
source is readily apparent at this position.

The pixel coordinates of this pre-explosion source are measured
to be $(1657.612\pm0.094, 2014.856\pm 0.048)$ using the average
results of three different centering algorithms in {\tt IRAF PHOT}.
This is offset from our estimated position of the SN 
by $(0.435,0.077)$ ACS pixels, or 0\farcs022 in total. Thus,
the SN and our progenitor candidate have formally coincident
positions given their respective uncertainties.  However,
this source also appears to be an extended blend of several
stars, with a FWHM of 3 pixels instead of the $\sim 2$ pixels
found for nearby point sources.  This proves to be a considerable 
complication for our photometric measurements. 

For photometry, we used {\sc hstphot} for the WFPC2 images and 
{\sc dolphot} for the ACS and WFC3 images (\citealt{Dolphin2000}).  {\sc hstphot}
is designed specifically for point-spread function (PSF)-fitting photometry 
on WFPC2 images, while {\sc dolphot} is a more general version of 
{\sc hstphot} which can also handle ACS and WFC3 data.
ACS, WFC3 and WFPC2 all have different pixel scales, and the
observations summarized in Table~\ref{tab:hstphot} were taken
with a range of orientations and depths.  Hence, it is difficult to 
directly compare observations taken with each of these cameras
particularly when the decomposition of the blended sources is not
unique.  The photometric results are reported in
Table~\ref{tab:hstphot} along with any magnitudes for the 
source available from the Hubble Source Catalog (HSC, \citealt{Whitmore2016}) for
comparison.  All the reported magnitudes are in the Vega system,
with appropriate transformations from the AB magnitudes used by the HSC.

The WFPC2 images were obtained with the WF3 detector, which has a pixel scale
of 0\farcs1 that grossly under samples the PSF.  The 1994 images were 
very shallow and there was a slight ($\sim 2$~pix) offset between the
two exposures, so we ran {\sc hstphot} on each frame individually. For
the WFPC2 data from 2001, there was no dithering between the exposures, so
we combined the two images available for each filter. Cosmic rays were
rejected using the paired exposures for each pointing.
{\sc hstphot} was run using its recommended parameters, performing PSF-fitting 
photometry with a detection threshold of 3.5$\sigma$, and simultaneously 
refitting the sky background (option 512). Aperture corrections were 
derived from the data, and applied to the measured magnitudes along with 
standard CTE corrections.  

\begin{figure}
\begin{center}
\includegraphics[width=4.0cm]{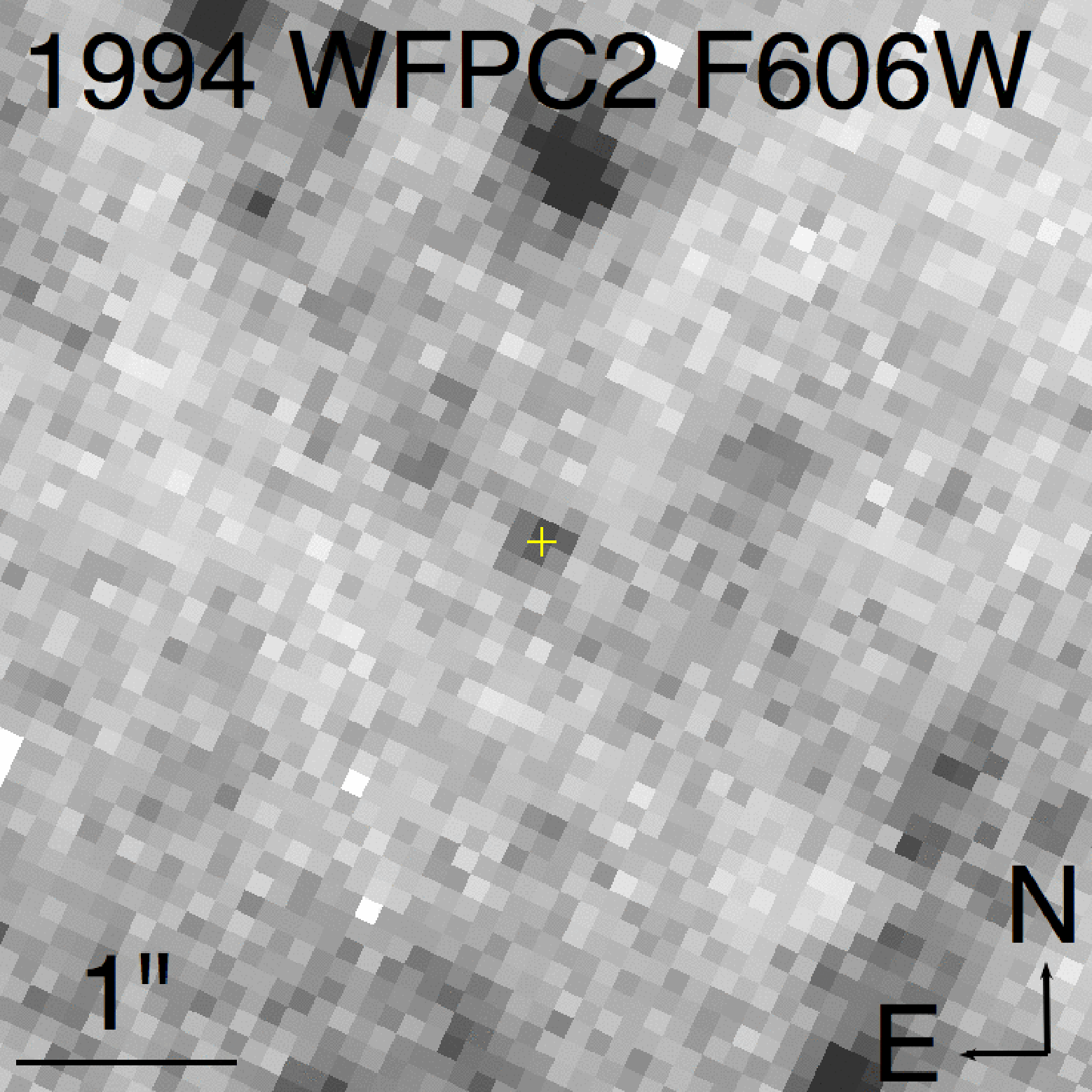}
\includegraphics[width=4.0cm]{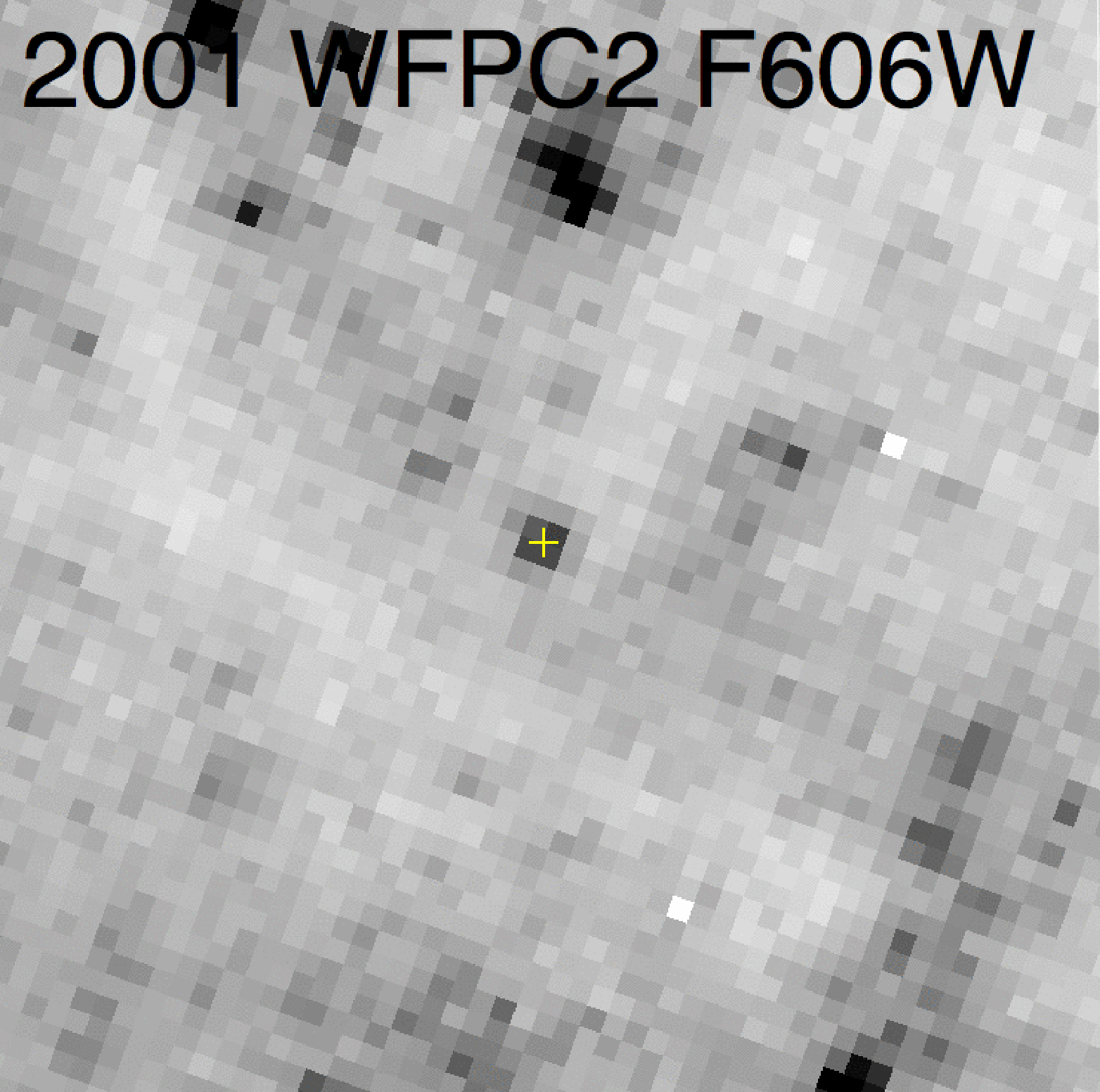}
\includegraphics[width=4.0cm]{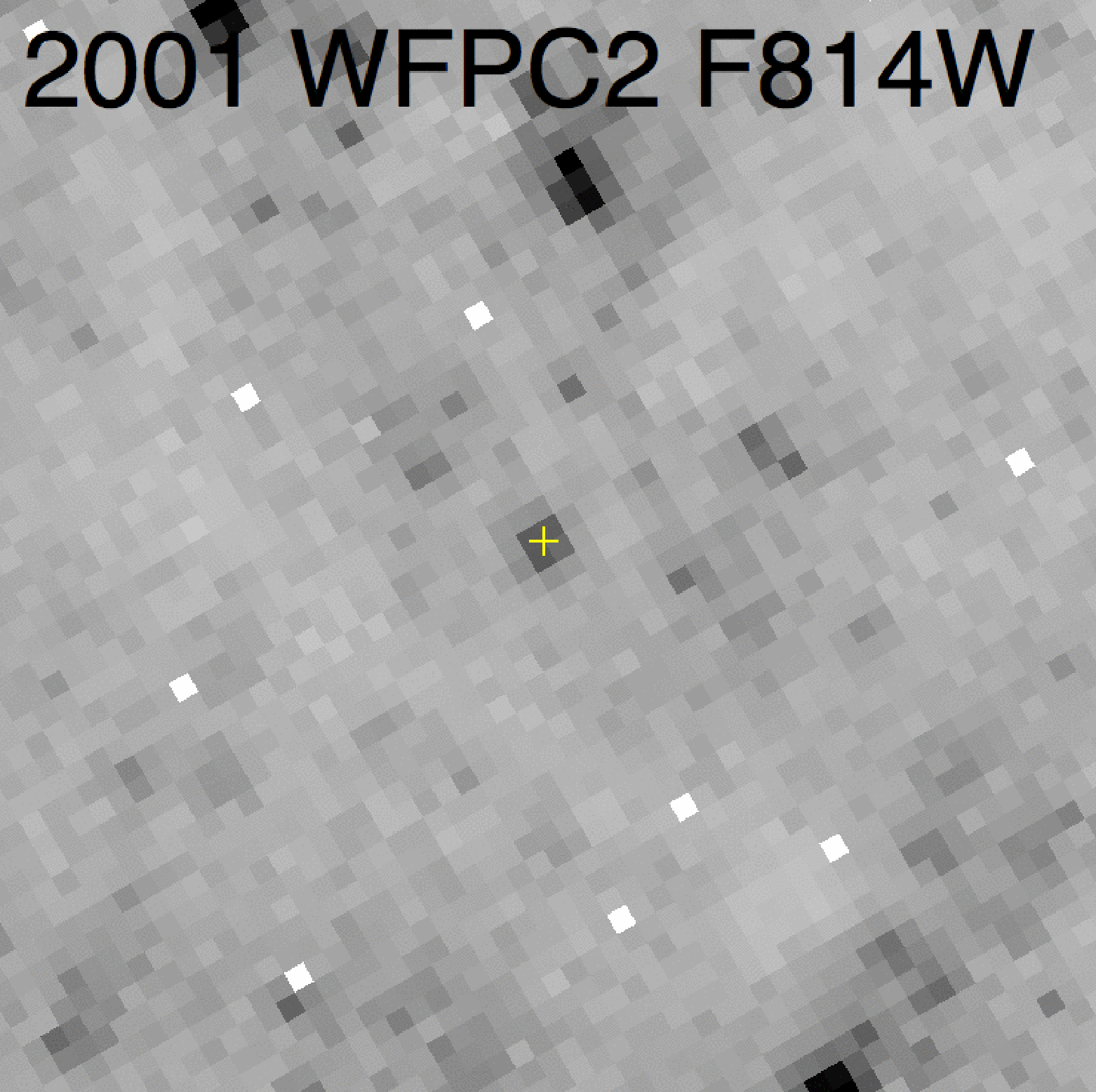}
\includegraphics[width=4.0cm]{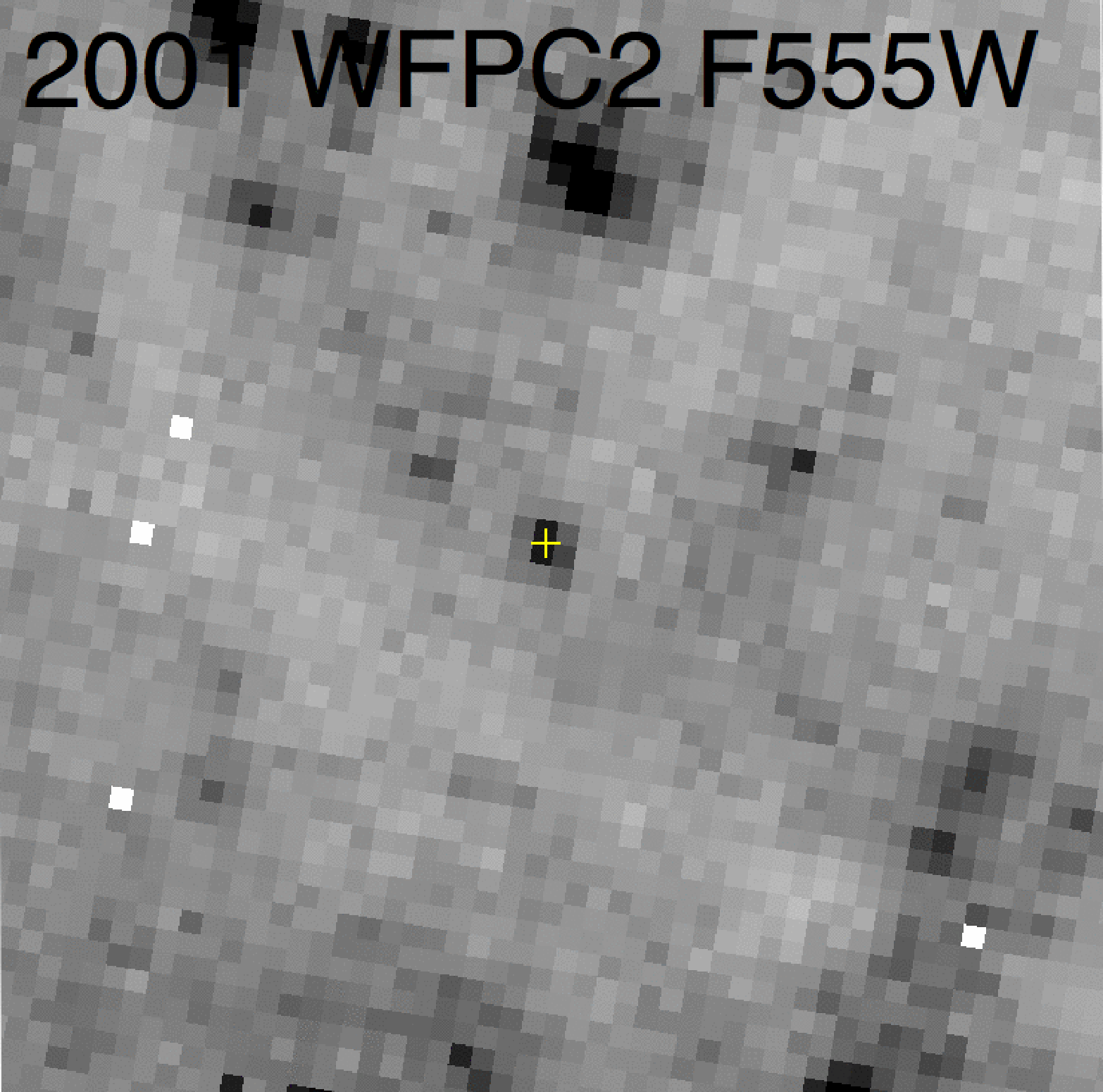}
\end{center}
\caption{ WFPC2 images ($5\farcs0$ square, North up) from 2001 centered on the candidate progenitor.
  The progenitor candidate is marked by the yellow cross.  The white squares are bad pixels.
  }
\label{fig:wfpc2}
\end{figure}

In all of the WFPC2 images, the progenitor candidate was detected as a 
single source, as shown in Fig. \ref{fig:wfpc2}.  The 1994 and 2001
F606W magnitudes differ by almost $0.5$~mag despite using identical
settings for {\sc hstphot}.  If we take 5 nearby
sources with $\hbox{F606W} \sim 22$~mag, we find that the sources
all appear to be brighter in 2001 by $0.15$ to $0.35$~mag.  
The cause of the discrepancy is unclear, although the 1994 exposures
were significantly shallower and have lower backgrounds that would
worsen the effects of CTE.  We applied the {\sc hotpants} difference
imaging package to the two F606W epochs and found no significant
residuals, indicating that there was no significant variability associated
with the source.  We will not consider the 1994 data further.  The
HSC had an estimate for the source flux in the 2001 F814W image,    
and the HST magnitude was consistent with our photometry 
to $<$0.1 mag.

\begin{figure*}
\begin{center}
\includegraphics[width=7.7cm,height=7.6cm]{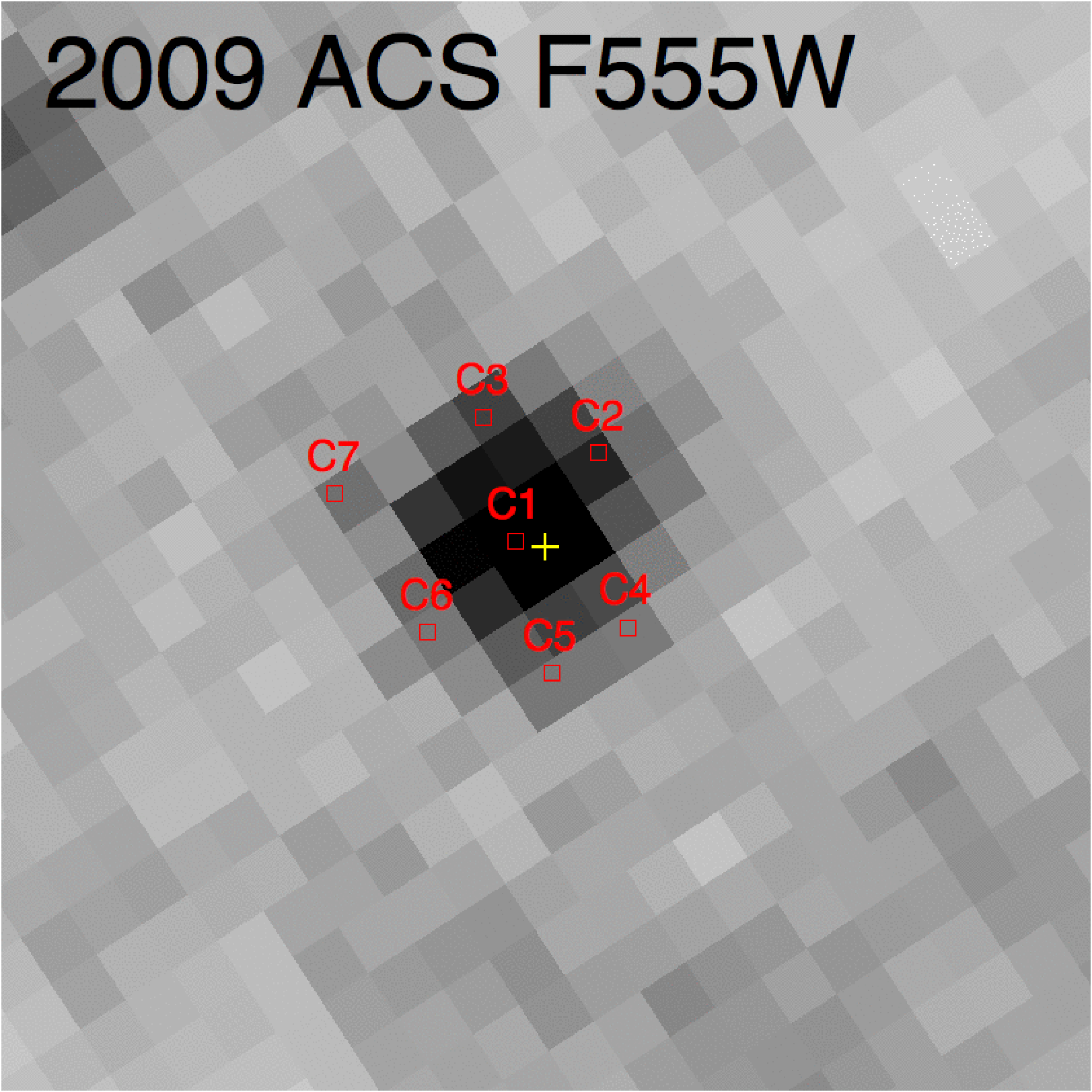}
\includegraphics[width=7.7cm,height=7.6cm]{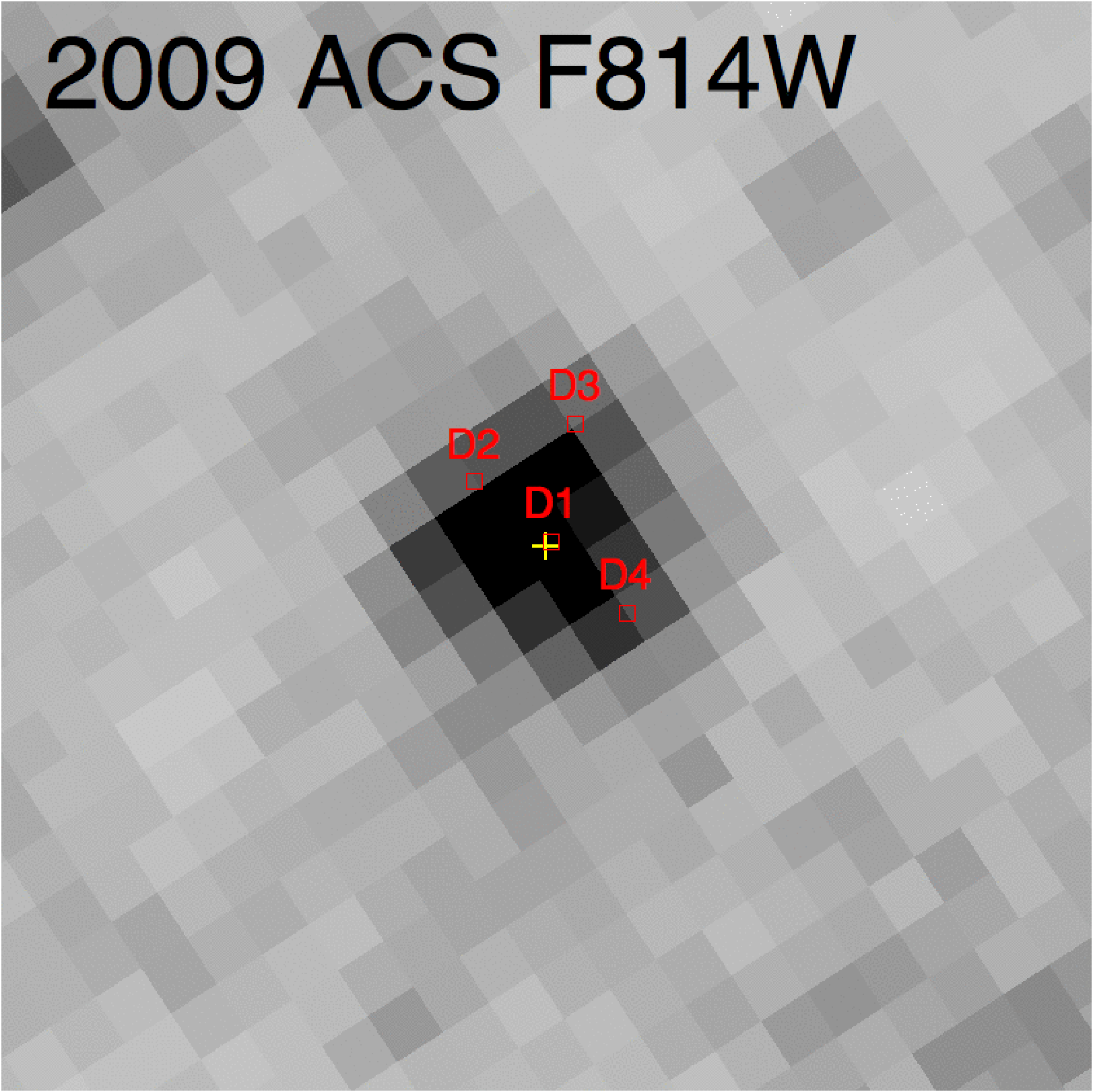}
\end{center}
\caption{
  ACS images (1\farcs0 square, North up) from 2009 (F555W and F814W, bottom)
  centered on the estimated position of ASASSN-16fq (yellow cross). 
  The labeled red squares show the positions of {\sc dolphot} sources reported in Table~\ref{tab:hstphot}
  where the central source (C1-D1) is the brightest. 
  }
\label{fig:acs}
\end{figure*}

The ACS images were analyzed with {\sc dolphot}.  The drizzled ACS 
reference image (file type { \tt \_drc}) for each filter was used to 
identify sources, while photometric measurements were performed on the 
individual, undrizzled, { \tt \_flc} files. Both file types are
corrected for CTE losses, so no CTE corrections were applied to the 
measured magnitudes.  The pipeline-reduced 2004 ACS { \tt \_flc} files 
did not have a cosmic ray mask in their data quality extension.  We
processed these files with {\sc astrodrizzle} to derive a cosmic ray mask
before carrying out the remainder of the analysis.  The {\sc dolphot}
parameters, including the choice of aperture radius, were matched to those
of the ANGST survey (\citealt{Dalcanton09}).  We used a large 8 pixel
aperture, fitting each source, its neighbors and the background
simultaneously.  We used a $3\sigma$ source detection threshold
and set the { \tt force=1} parameter to force all objects to be
fit as point sources.  For the F658N image,  {\sc dolphot} failed to 
detect enough sources to align the two { \tt \_flt} images to the 
drizzled {\it F658N} reference image. Here we measured the positions of 
many point-like sources common to each image within {\sc iraf}, 
and used {\sc acsfitdistort} within {\sc dolphot} to align the 
frames.  The location of the SN in each of the images was found by
aligning them to the 2009 F814W image which was used to identify the progenitor.

{\sc dolphot} splits the flux from the source at the progenitor position 
into a number of sources, which is not surprising given that that its
FWHM is broader than other nearby, point-like sources.  If we disable
the { \tt force=1} option, the source is modeled as a single extended
object rather than decomposed into multiple sources.  Figure~\ref{fig:acs} 
shows the relevant region of each ACS image and how it has been
decomposed into the sources reported in Table~\ref{tab:hstphot}.
The HSC magnitudes agree well ($\sim 0.1$~mag) with our results for 
the F435W, F555W and F658N filters.  The difference is much larger
for F814W, where {\sc dolphot} has found two sources (D1 and D2) of
similar flux.  The summed flux of these two sources, corresponding
to $22.74$~mag, agrees with the HSC flux.

\begin{figure}
\begin{center}
\includegraphics[width=8cm]{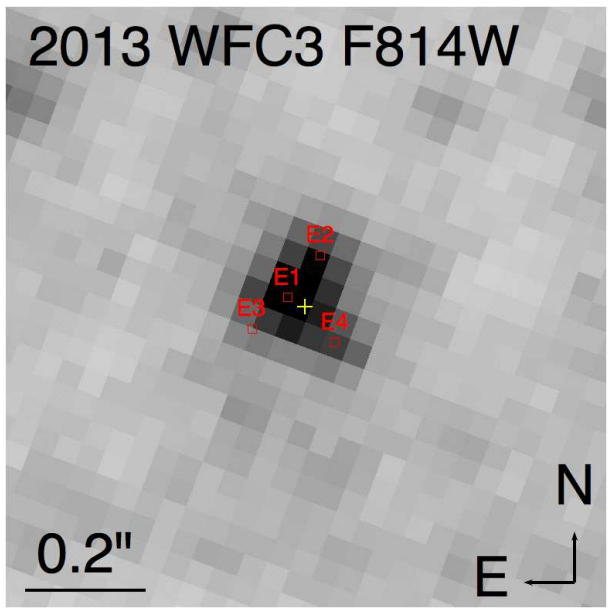}
\caption{ WFC3 F814W image (1\farcs0 square, North up) from 2013 centered on the
  estimated position of ASASSN-16fq (yellow cross).
  The labeled red squares show the positions of {\sc dolphot} sources reported in Table~\ref{tab:hstphot}
  where the central source (E1) is the brightest.
   }
\label{fig:wfc3}
\end{center}
\end{figure}

We also analyzed the WFC3 images with {\sc dolphot}, where these are the images
that best sample the PSF (0\farcs04 pixels rather than the 0\farcs05 scale of ACS) and likely 
produce the most reliable source decomposition.  We identified the sources on 
the {\it F814W} image and then obtained photometry for the {\it F555W} and {\it F814W} 
images simultaneously.  The sources identified by {\sc dolphot} are shown in 
Figure~\ref{fig:wfc3} and their measured magnitudes are reported in Table~\ref{tab:hstphot}. 
The position of ASASSN-16fq was again determined by aligning the WFC3 F814W image to
the ACS F814W image. 

The photometry reported in Table~\ref{tab:hstphot} is not fully consistent given
the reported uncertainties.  This is not very surprising given that the source
appears to be a blend of multiple sources with a decomposition that is not unique
under changes in the instrument, camera and filter.  We tested running {\sc 
dolphot} simultaneously on the ACS and WFC3 F814W images, using a single 
drizzled WFC3 image for source detection.  We found systematic offsets of
$0.1$-$0.2$~mag between sources as measured on the ACS and WFC3 images.  For
isolated point sources, the differences should be much smaller (a few times
$0.01$~mag in tests with {\sc synphot}).  However, even with the source 
positions fixed, the flux estimates for the blended sources likely depend
in detail on the sub-pixel scale model of the PSF in each frame.

\begin{figure*}
\begin{center}
\centerline{
  \includegraphics[width=5.5cm]{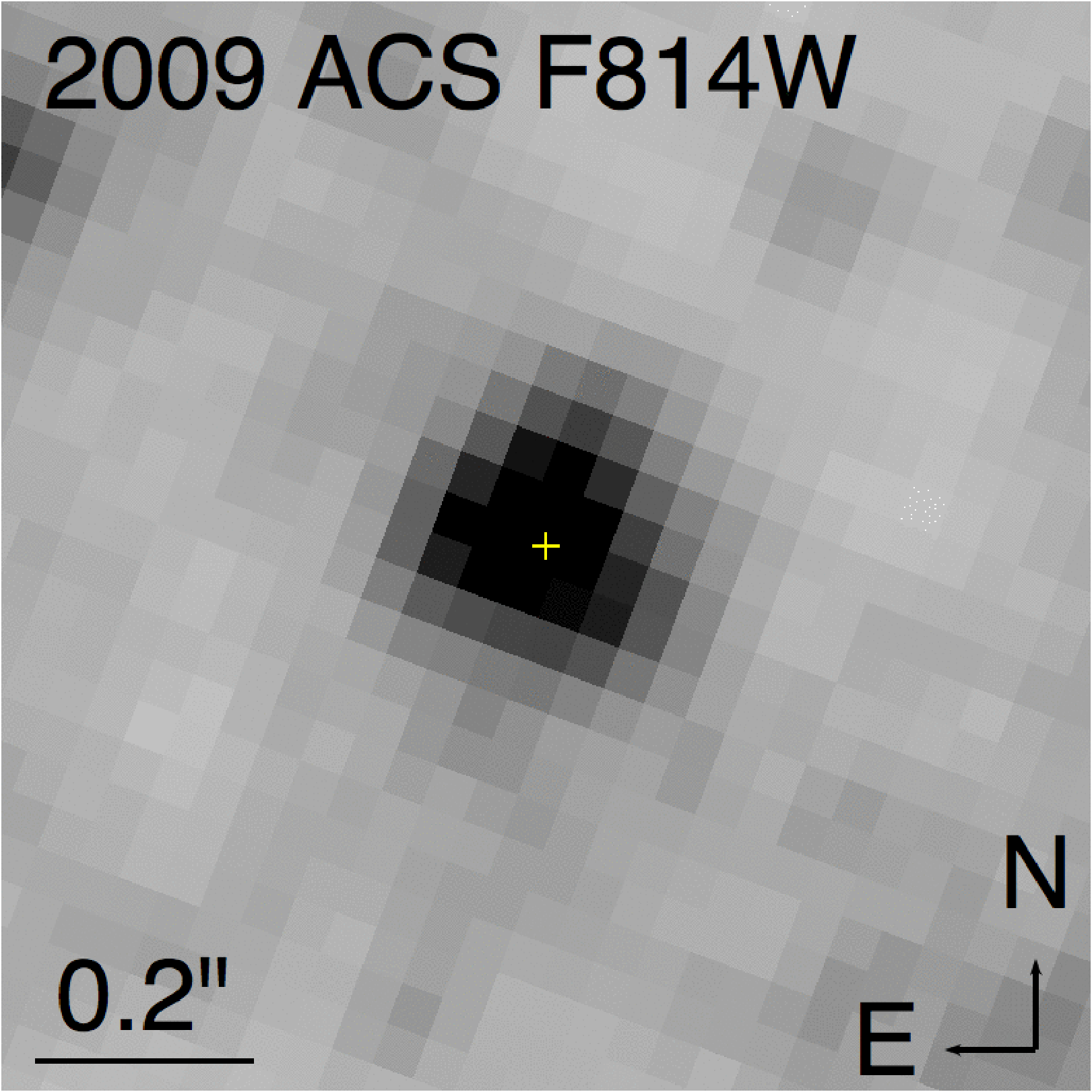}
  \includegraphics[width=5.5cm]{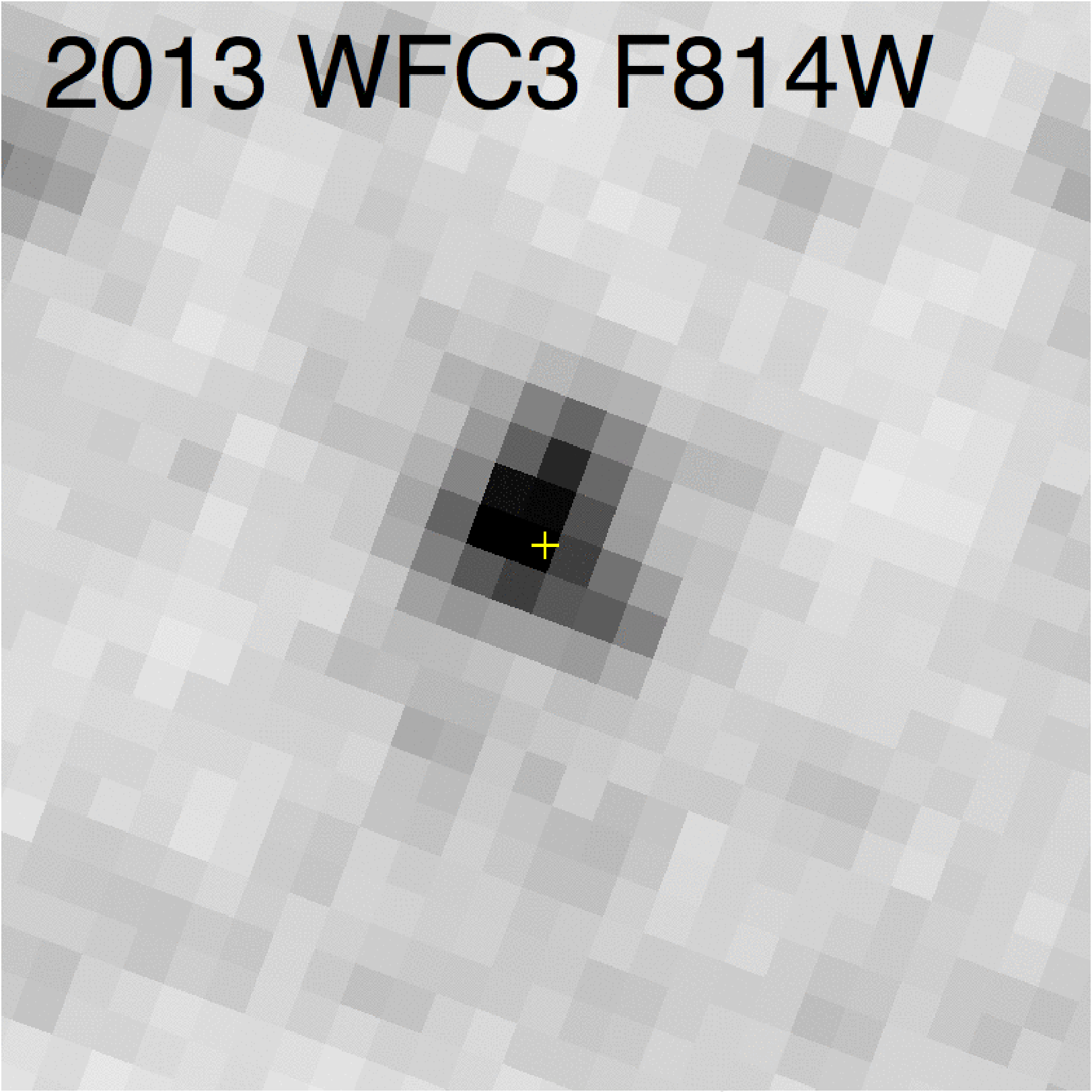}
  \includegraphics[width=5.5cm]{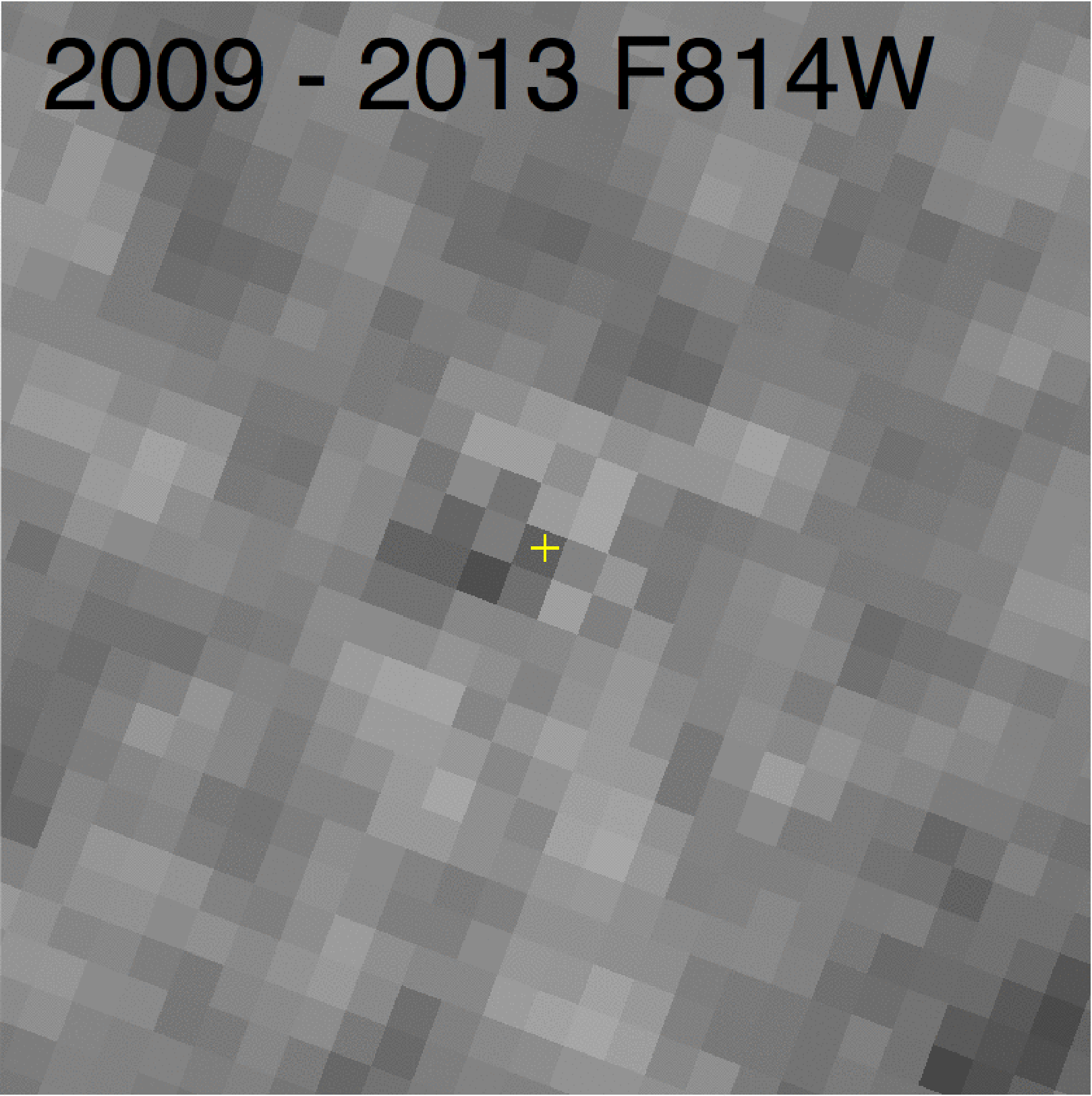}
  }
\caption{ No evidence for variability at F814W.  The 1\farcs0 square (North up)
  panels show the ACS (left) and WFC3 (middle)
  F814W images from 2009 and 2013, respectively, along
  with the {\sc hotpants} difference image (right) between them.  The position of
  the progenitor candidate for ASASSN-16fq is marked with a yellow cross in all three panels.
  In the difference image, sources which are brighter in the ACS image appear dark while those
  that are brighter in the WFC3 image appear white.
   }
\label{fig:f814w_sub}
\end{center}
\end{figure*}

We again used difference imaging to search for evidence that the differences
between the ACS and WFC3 magnitudes could be explained by variability.  We
focused on the F814W filter, where a red supergiant exploding as a Type~II
SN should contribute the most flux.  As shown in Figure~\ref{fig:f814w_sub},
there are no significant residuals after using {\sc hotpants} to scale and
subtract the two images.  As with the similar test on the WFCP2 F606W images
above, there is no evidence in the difference image for any variability.

We also searched the archival Spitzer data for sources related to the
progenitor following the procedures of \cite{Khan2015b}. 
We used the {\sc ISIS} (\citealt{Alard1998}) difference imaging package
to align the Spitzer images of a sub-region centered on ASASSN-16fq.
Many of the images had an artifact passing close to the SN, so we
combined the epochs either missing the artifact or where the artifact
avoids its location to build a reference image.  We then produced
differenced images to search for variability.  We also differenced
the $3.5$ and $4.5\mu$m reference images to search for dusty stars.
In such a ``wavelength difference'' image, the normal stars all
vanish to leave only stars with significant hot dust emission  
because they all have a common ``Rayleigh-Jeans'' spectral
energy distribution (SED) (\citealt{Khan2010}).  We found no evidence for
mid-IR variability or hot dust emission.  There is no mid-IR 
source at the position of the progenitor. Our best estimates of the
$3\sigma$ upper limits on the $3.6$, $4.5$, $5.8$ and $8.0\mu$m 
fluxes of any source at the position of the progenitor are  $15.9$, $15.7$, $13.3$ and 
$11.4$~mag, respectively.   The shorter wavelength limits correspond
to $\nu L_\nu \ltorder 10^5 L_\odot$ (at $1\sigma$),  and provide
no strong constraints given the HST photometry.

Clearly with the source blending and the differences between the 
results for the various cameras we cannot precisely determine the
properties of the progenitor or the amount of extinction.  
Our goal is simply to provide constraints on the progenitor 
luminosity that can be translated into rough constraints on 
the stellar mass.  
ASASSN-16fq is a Type~IIP SN, and all progenitors of Type~IIP SN with 
constrained stellar temperatures have been red supergiants (see
\citealt{Smartt2015}) with stellar temperatures of order
$T_*=3500$~K.  Higher temperatures of order $T_*=6000$~K have
been observed for the progenitors of Type~IIb SNe (e.g., SN~2011dh, 
\citealt{Maund2011}).  Finally, we must note the still higher
temperature progenitor of the Type~IIpec SN~1987A, a blue supergiant with
$T_* \simeq 16000$~K and $L_* \simeq 10^{5.0} L_\odot$ (see the review by
\citealt{Arnett1989}).  All available evidence is that ASASSN-16fq
is a normal Type~IIP SN where we should only consider the lowest
temperature ($T_*=3500$~K), but we will present parallel results
for $T_*=6000$~K and comment on the consequences of still higher
progenitor temperatures.   

Rodriguez et al. (2016, in preparation) 
modeled the early, near-UV/optical/near-IR photometry
(from {\it Swift}, LCOGT~1m, the Iowa Observatory~0.5m, and
the REM~0.6m) and photospheric expansion velocities of ASASSN-16fq
(obtained from optical spectra obtained with FLWO~1.5m+FAST and
MDM~2.4m+OSMOS) following \cite{Pejcha2015b}. This
phenomenological model describes the multicolor light curves and
photospheric velocity evolution of normal Type~II SNe, decoupling
changes in effective temperature from changes in the photospheric
radius at different epochs since explosion. These fits lead
to an estimated explosion time of $\rm JD=2457532.2 \pm 2.0$~days, a total
extinction (including Galactic) of $\rm E(B-V)=0.53 \pm 0.02$~mag,
and a distance modulus of $29.72 \pm 0.44$~mag. This distance
estimate of $8.8 \pm 1.8$~Mpc is consistent with our adopted
value of $10.62$~Mpc.  For roughly estimating the properties
of the progenitor, we will assume either Galactic 
$\rm E(B-V)= 0.03$ or $\rm E(B-V)= 0.53$ of foreground 
extinction.  The results assuming circumstellar
extinction would be moderately different (see \citealt{Kochanek2012}
for a discussion of this issue).
  
\begin{figure*}
\centering
\includegraphics[width=0.45\textwidth]{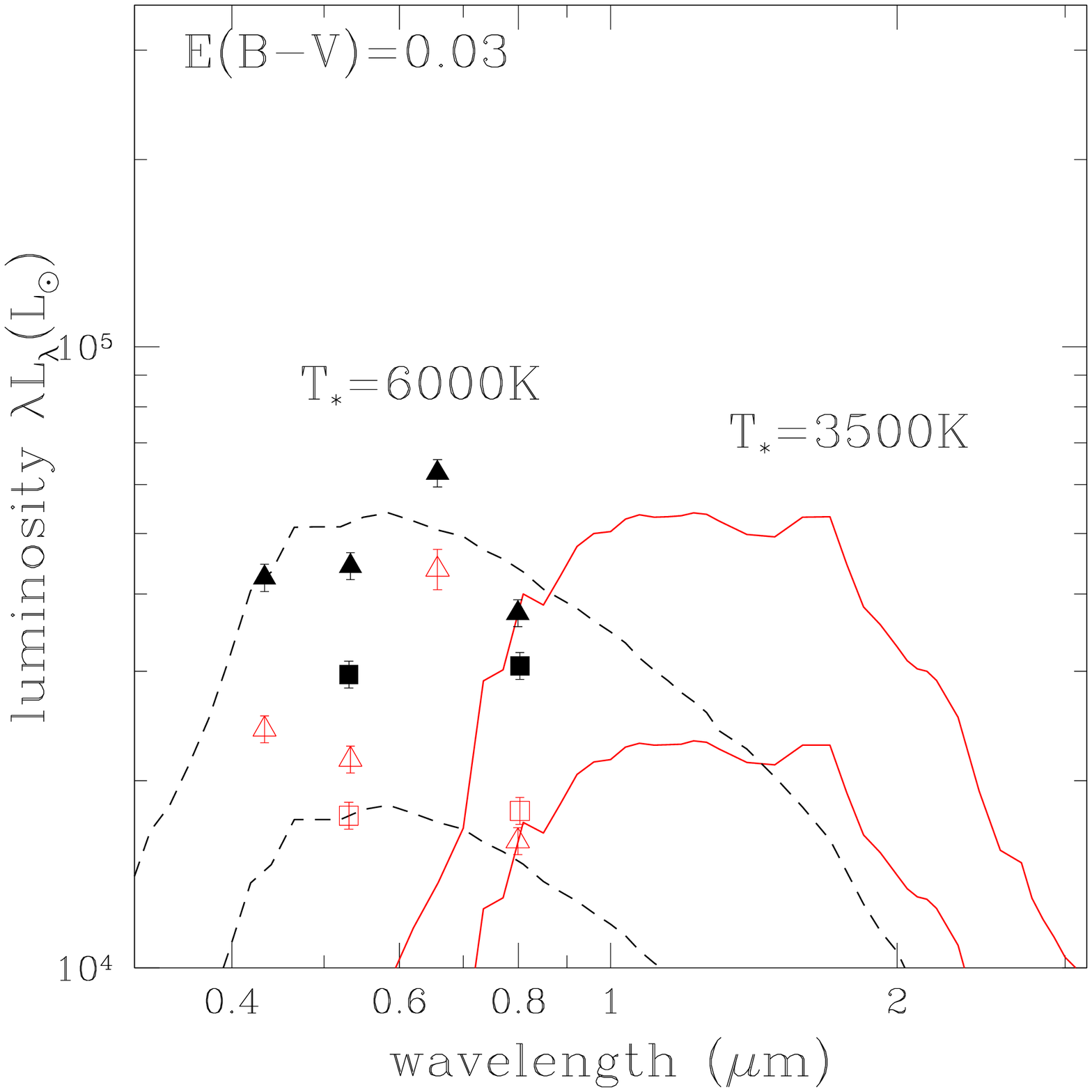}
\includegraphics[width=0.45\textwidth]{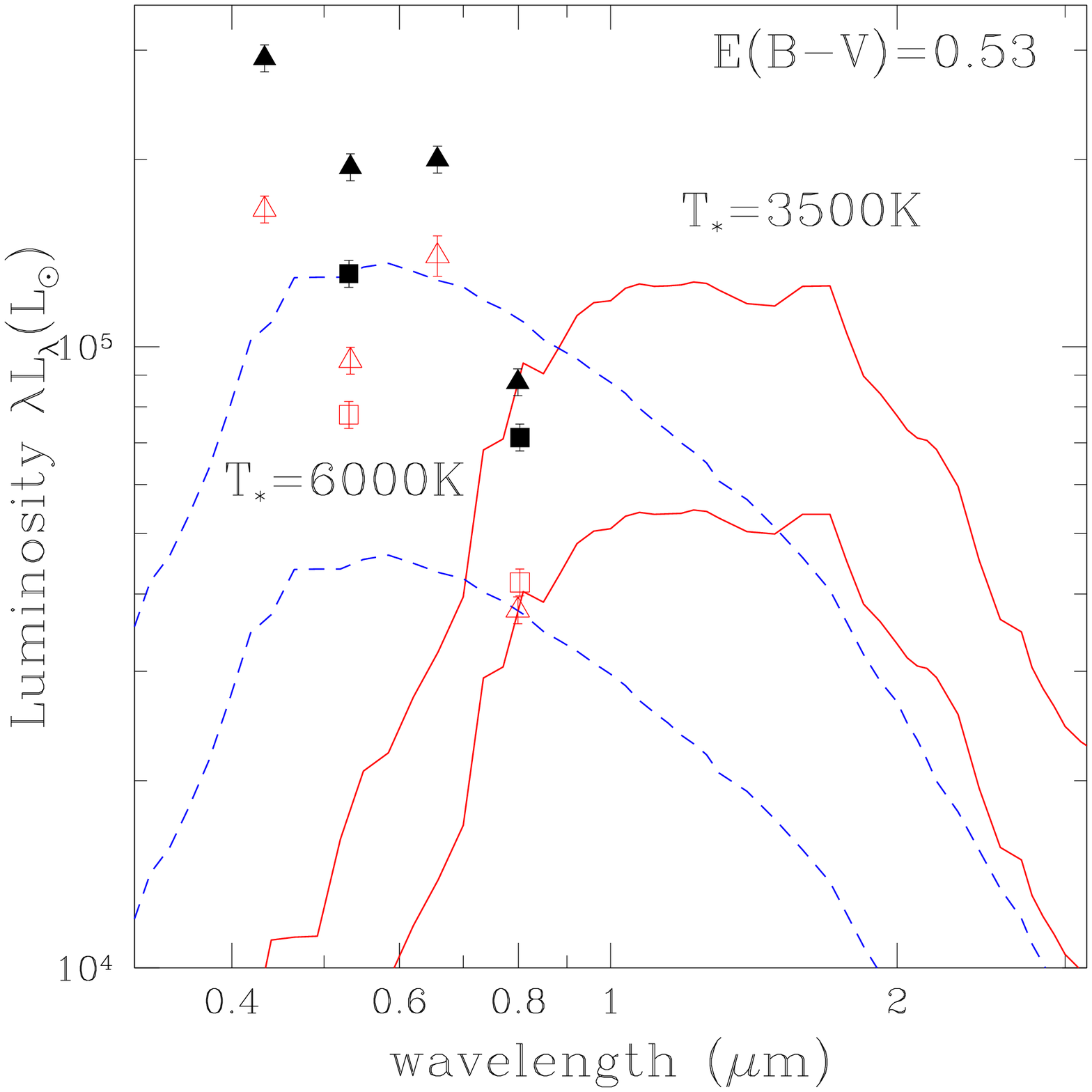}
\caption{Spectral energy distributions with low ($\rm E(B-V)=0.03$, left) and 
high ($\rm E(B-V)=0.53$, right) extinction.  The open (filled) squares show the WFC3
photometry for the brightest source (total of all sources) in Table~\ref{tab:hstphot}.
The open (filled) triangles show the ACS photometry for the brightest
source (total of all sources) in Table~\ref{tab:hstphot}.  The solid
red (dashed blue) curves show the $T_*=3500$~K ($T_*=6000$~K) stellar
SEDs with the lowest and highest luminosities that fit one of the
broad band photometric points and do not significantly exceed the
brightest, total broad band fluxes (the total ACS fluxes for the
broad band filters). 
}
\label{fig:sed}
\end{figure*}

We corrected the photometry for the assumed
amount of foreground extinction and then used the {\sc DUSTY} dust 
radiation transport code (\citealt{Ivezic1997}, \citealt{Ivezic1999}, 
\citealt{Elitzur2001}) combined with \cite{Castelli2004} model 
atmospheres of solar metallicity and embedded in a Monte Carlo
Markov Chain (MCMC) driver to match the model to the data
(see \citealt{Adams2015}, \citealt{Adams2016} for details).
For the experiments we carry out here, much of this machinery
is not necessary but can be trivially applied to the problem.
For a fixed stellar temperature and foreground extinction we
determined the stellar luminosity based on one of the broad
band ACS or WFC3 total or brightest component magnitudes. The
brightest, total broad band magnitudes (those from ACS) were
used as $1\sigma$ upper limits on the luminosity unless 
the magnitude was being used to determine the luminosity.
In essence, we will not attempt to extract any color information
from the photometry, but no model can be significantly brighter
than the brightest observed ACS/WFC3 broad band magnitudes.
   
Figure~\ref{fig:sed} shows the ``SED'' of the source combining
the ACS and WFC3 photometry.
All else being equal and ignoring the narrow band
F658N photometry, the luminosity estimate will vary by roughly
a factor of two between a model based on the fluxes of the
brightest components (A1-E1) and the summed fluxes of all 
the components.  For example, using $T_*=3500$~K and 
Galactic extinction, only models normalized at I-band (F814W)
are also consistent with the relevant upper limits.  They
have luminosities in the range from $L_* = 10^{4.5}$ to
$10^{4.9}L_\odot$.  Essentially, the cool models are too
luminous for the observed F814W fluxes if they are normalized
at shorter wavelengths.  For the hotter $T_*=6000$~K model, 
all permutations are allowed, and they have luminosities
spanning a very similar range from $L_*=10^{4.5}$ to $10^{4.8}L_\odot$.   
At the hotter temperature, the stellar SED is not dropping 
rapidly in the bluer bands and so it is easier to be 
consistent with all the constraints.
When we raise the extinction to $\rm E(B-V)=0.53$, the
changes in the results are modest.  For $T_*=3500$~K,
only the F814W normalizations are again allowed but
with a shift to moderately higher luminosities of
$10^{4.8}$ to $10^{5.2}L_\odot$.  For $T_*=6000$~K
the F814W and F555W normalizations are allowed and
they have the same luminosity range as for the
$T_*=3500$~K models.  

A hot star, albeit one still 
hotter than the progenitor of SN~1987A, might also explain
the discrepant F658N magnitudes as H$\alpha$ emission 
from an associated HII region.  With the large 
$\rm E(B-V)=0.53$ extinction correction, the implied
SED also rises rapidly to shorter wavelengths as might
be expected for a hot star.  Since the SED rises rapidly
to shorter wavelengths, these hot star models are very
luminous, with $L_* = 10^{5.5}$  to $10^{6.0}L_\odot$
for $T_*=16000$~K  and still higher luminosities for
higher temperatures.  We will not consider this 
possibility further, although it would make ASASSN-16fq
a unique and fascinating new example of the SN
phenomenon were it to prove to be true.

If we use the endpoints of the PARSEC (\citealt{Bressan2012})
isochrones to map luminosities into masses, luminosities of
$L_*=10^{4.5}$, $10^{4.8}$, $10^{4.9}$ and $10^{5.2}L_\odot$
translate into initial masses of $M_* \simeq 8.6$, $11.2$,
$11.6$ and $17 M_\odot$.  Since the higher luminosity 
limits correspond to the cases normalized by the total
flux of all the detected sources rather than just that of
the brightest source found by {\sc dolphot}, they should
probably be treated as upper limits.  In short, despite 
the chaotic nature of the photometry, the progenitor of
ASASSN-16fq was probably a red supergiant with an initial
mass of $8$-$12M_\odot$.  This estimate of the mass range
is fairly robust to changes in temperature, unless the star
is significantly hotter than expected ($T_* > 6000$~K), and 
holds for a fairly broad range of extinctions.  It probably does 
require that some of the blended stars are significantly
hotter and bluer, which would be natural if they are also
younger.    

\section{LBT and the Variability of the Progenitor}

As part of the LBT search for failed SN (\citealt{Kochanek2008},
\citealt{Gerke2015}), NGC~3627 was observed
sixteen times in the U, B, V and R-bands between 4 May 2008 and
7 February 2016.  NGC~3627 lies on chip 3 of the Large 
Binocular Camera (LBC, \citealt{Giallongo2008}) images. The LBT data are analyzed
using the ISIS (\citealt{Alard1998}) difference imaging package.  First, all
the images are aligned to a common R-band astrometric reference
image.  The best images (noise, resolution and
quality) for each band are combined to make a reference 
image.  These reference images are astrometrically and photometrically 
calibrated based on SDSS stars (\citealt{Pier2003}, \citealt{Ivezic2007}, 
\citealt{Ahn2012}) with the SDSS ugriz photometry transformed to UBVR 
Vega magnitudes using the relations in \cite{Jordi2006}.
The reference images are then subtracted from the individual epochs 
to leave only the variable flux between the reference image and the individual
epochs.  Since the reference image is constructed from images
spanning the monitoring period, source fluxes in the reference image
roughly correspond to the mean fluxes.

We cannot measure the absolute flux of the progenitor in the
LBT images due to crowding.  However, the subtracted images
are not crowded and we can easily constrain the variability
of the progenitor.  For this purpose, the uniformity of the data and the
well-sampled PSF makes the ground-based LBT data superior to the
archival HST data. Although variability is seen in none of the bands,
we focus only on the R-band data, which is both deeper and of 
higher quality.  Table~\ref{tab:lbtphot} provides the
difference in the R-band luminosity relative to the 
error-weighted mean of all the epochs.  The conversion
between the R-band ($\nu L_\nu$) luminosity and counts 
is $3.26 L_\odot$/count and no correction for extinction 
has been applied.

\begin{figure}
\centerline{\includegraphics[width=3.5in]{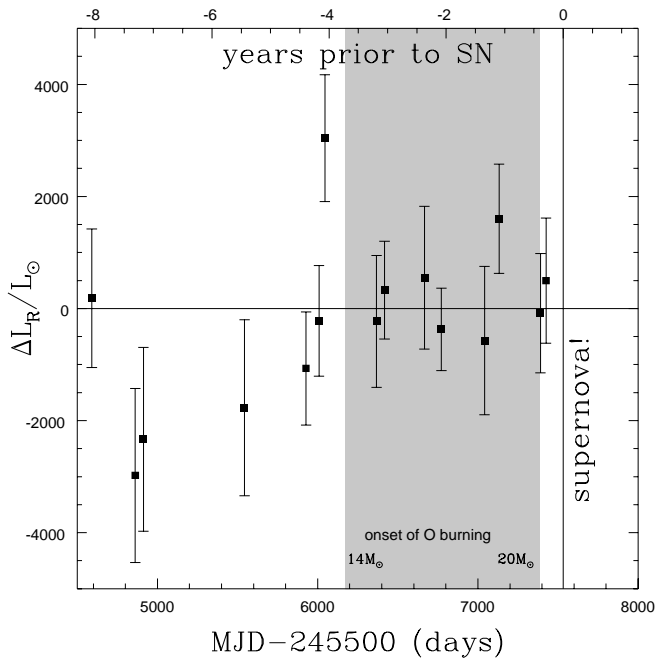}}
\caption{
  R-band progenitor variability prior to the supernova relative to
  the error-weighted mean.  The luminosities are not corrected for
  extinction.  The gray band shows the times for the commencement
  of core oxygen burning for stars with initial mass of $14M_\odot$
  (left edge) to $20 M_\odot$ (right edge) from 
  \protect\cite{Sukhbold2014} (see Figure~\protect\ref{fig:bigpicture}
  for the full mass dependence).
  }
\label{fig:vary1}
\end{figure}

\begin{figure*}
\centerline{\includegraphics[width=3.5in]{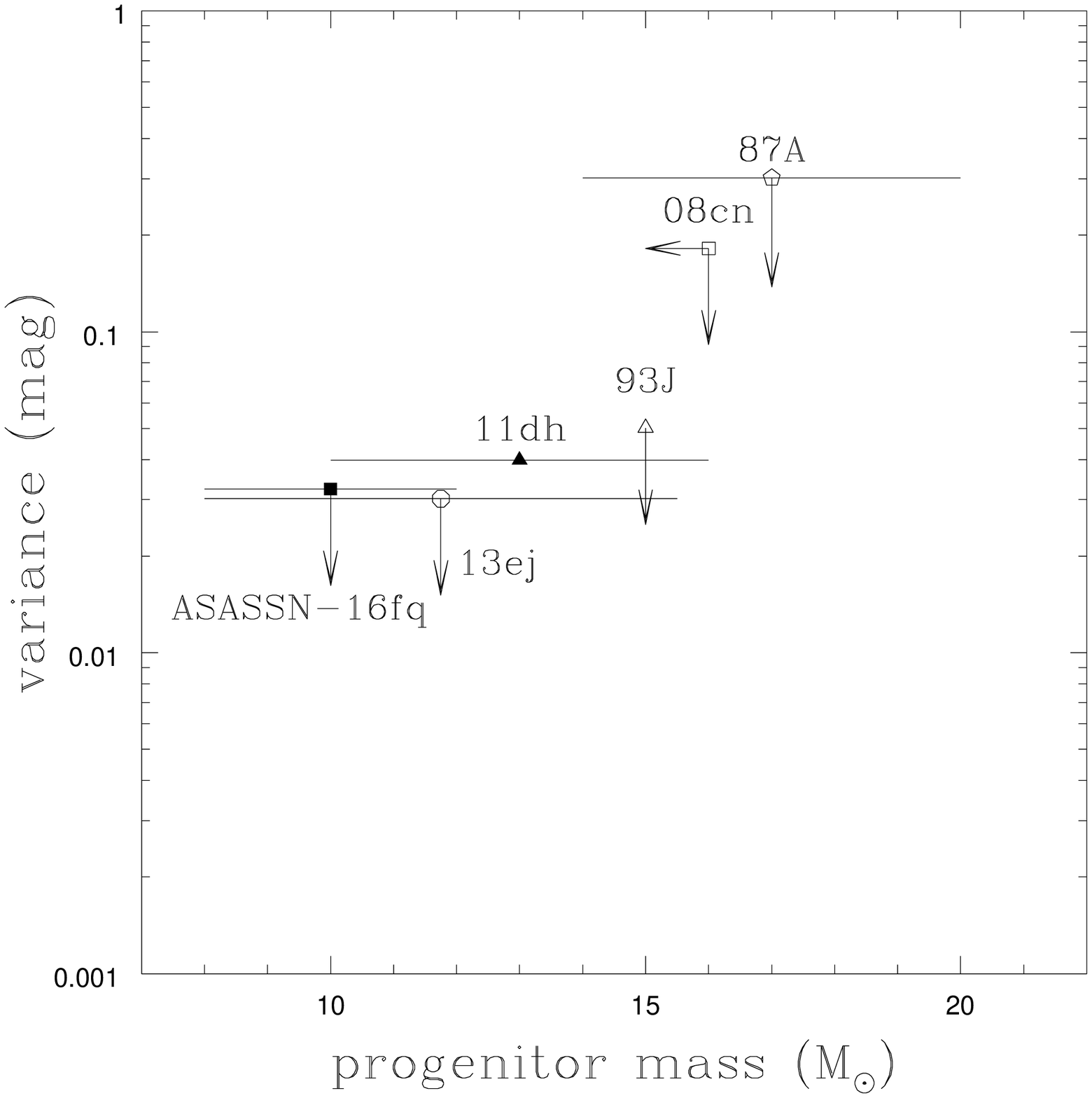}
            \includegraphics[width=3.5in]{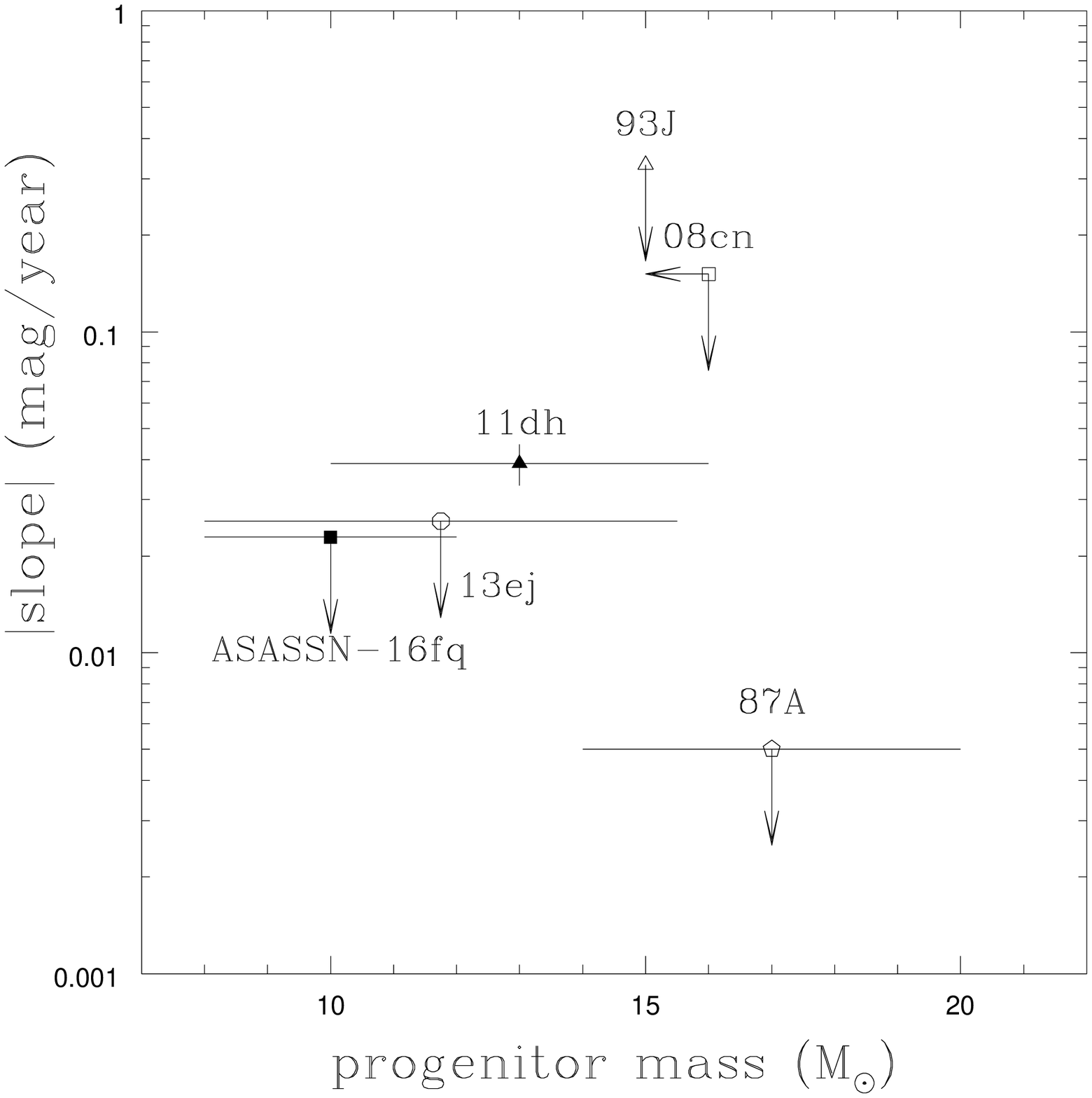}}
\caption{
  (Left) Magnitude variance of the 6 SN progenitors as a function of
  their estimated mass.  The variance in the progenitor's flux is a measure of
  any ``random'' variability of the progenitor.  
  The time period over which the variance was estimated varies between the SN
  (see Figure~\protect\ref{fig:bigpicture}).
  }
\label{fig:vary2}
\caption{
  (Right) Absolute values of the linear luminosity slopes of the 6 SN progenitors as a function of
  their estimated mass.  The slope of the progenitor's flux is a measure of any
  ``steady'' variability of the progenitor.  For the SN where we have only
  upper limits, the limit is drawn at the absolute value of the slope plus
  the uncertainty in the slope.
  The time period over which the slope was estimated varies between the SN
  (see Figure~\protect\ref{fig:bigpicture}).
  }
\label{fig:vary3}
\end{figure*}

\begin{figure*}
\centerline{\includegraphics[width=6.8in]{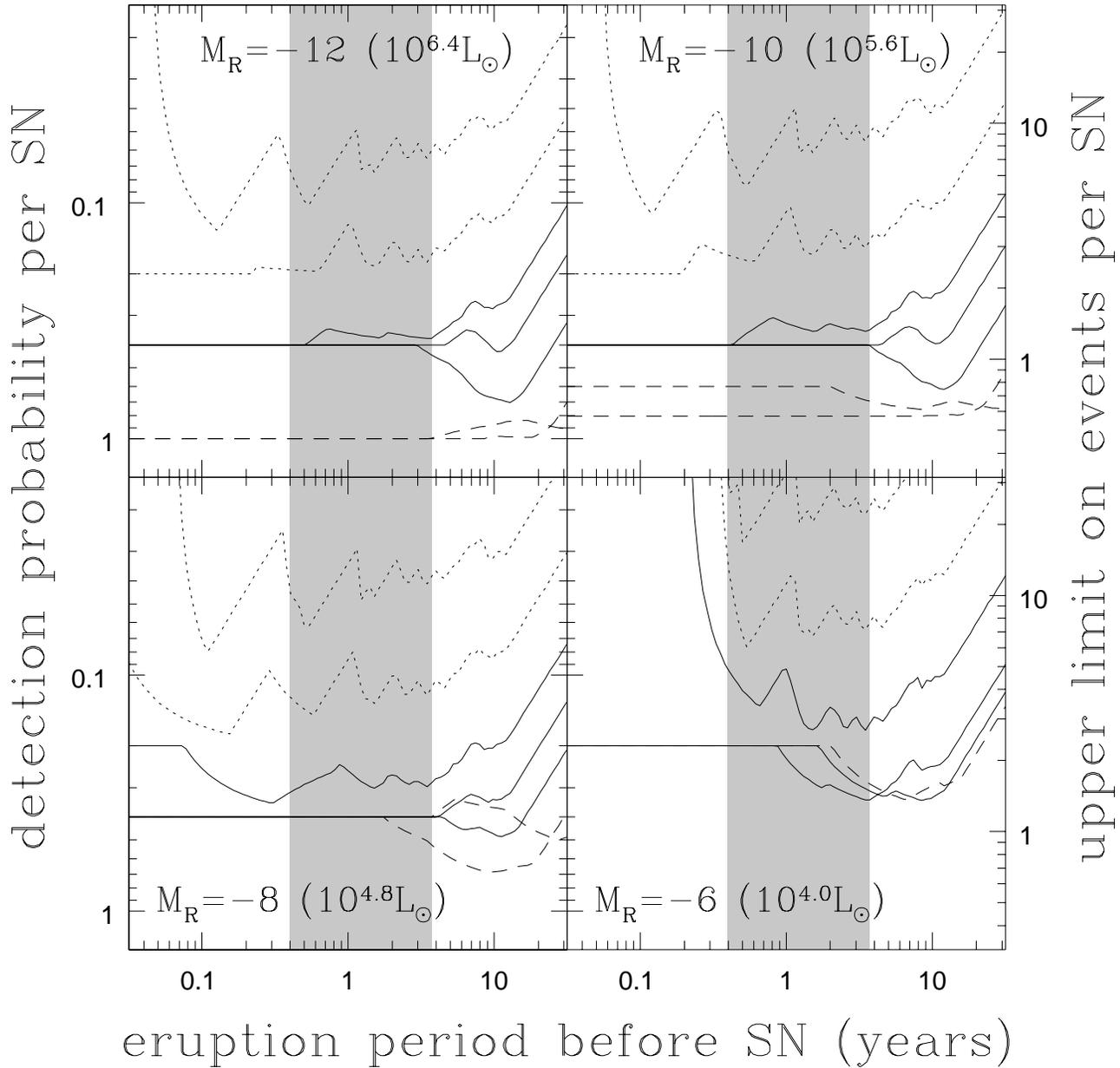}}
\caption{
  Outburst detection probabilities ($P_d$, left scale) and 90\% confidence upper
  limits on the number of outbursts per SN ($N_{out}$, right scale) as a function of
  the period $t_{out}$ over which eruptions can occur prior to the SN.  The panels show the results for
  different peak luminosities from $M_R = -12$ (top left) down to
  $M_R = -6$~mag (bottom right) where the associated luminosities are
  $\lambda L_\lambda$ at R-band.  
  The curves correspond to logarithmically
  spaced outburst FWHM of $t_{peak}= 0.032$ (highest, dotted), $0.1$
  (dotted), $0.32$ (solid), $1.0$ (solid), $3.2$ (solid), $10.0$ (dashed)
  and $32$~ (usually lowest, dashed) years.  
   As in Figure~\protect\ref{fig:vary1},
  the shaded region shows the time before collapse for the onset of
  core oxygen burning in the $14M_\odot$ and $20M_\odot$ models
  (see Figure~\protect\ref{fig:bigpicture}).
  The limit on the number of $N_{out}$
  outbursts and the detection probability $P_d$ are related by
  $N_{out}=2.3( N_{SN} P_d)^{-1}$ for the $N_{SN}=5$ SN excluding SN~1987A.
  }
\label{fig:outburst}
\end{figure*}

\begin{figure*}
\centerline{\includegraphics[width=6.8in]{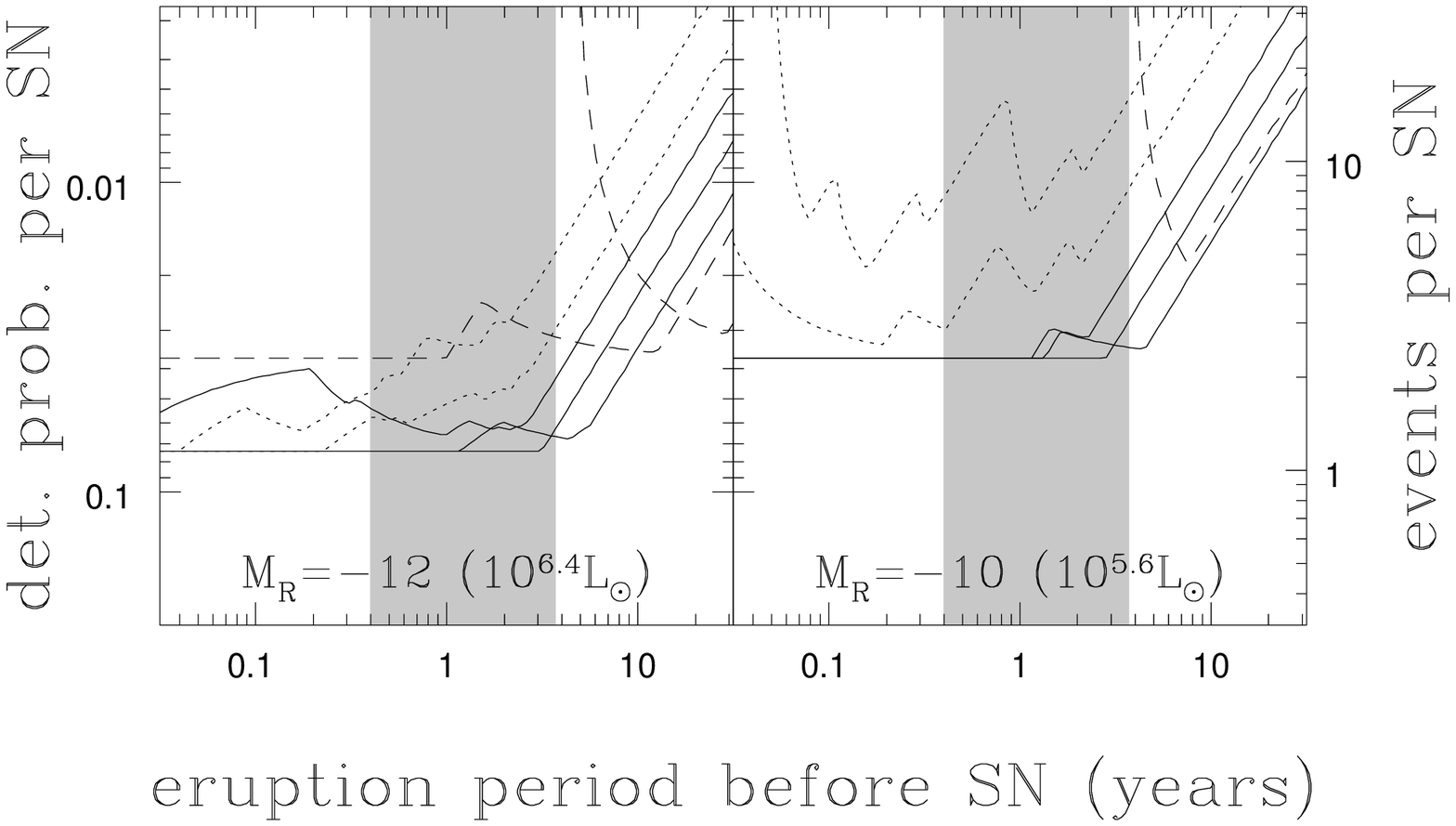}}
\caption{
  Outburst detection probabilities (left scale) and 90\% confidence upper
  limits on the number of outbursts per SN (right scale) as a function of
  the period over which eruptions can occur prior to the SN for the PTF
  data on Type~IIb SN progenitors considered by \protect\cite{Strotjohann2015}.
  The format is the same as in Figure~\protect\ref{fig:outburst}, but we only show
  the results for $M_R=-10$ and $-8$~mag where there is some overlap with
  the sensitivities shown in Figure~\protect\ref{fig:outburst}.  
  The curves correspond to logarithmically
  spaced outburst FWHM of $t_{peak}= 0.032$ (highest, dotted), $0.1$
  (dotted), $0.32$ (solid), $1.0$ (solid), $3.2$ (solid), $10.0$ (dashed)
  and $32$~ (usually lowest, dashed) years.  
  The PTF sample
  is better for brighter transients (more SN with better temporal coverage)
  and worse for fainter transients (insufficient depth).
   As in Figure~\protect\ref{fig:vary1},
  the shaded region shows the time before collapse for the onset of
  core oxygen burning in the $14M_\odot$ and $20M_\odot$ models
  (see Figure~\protect\ref{fig:bigpicture}).
  The limit on the
  number of $N_{out}$ outbursts and the detection probability $P_d$ are related by
  $N_{out}=2.3( N_{SN} P_d)^{-1}$ for the $N_{SN}=27$ SN in the
  \protect\cite{Strotjohann2015} sample.
  }
\label{fig:outburst2}
\end{figure*}

As seen in Figure~\ref{fig:vary1} and Table~\ref{tab:lbtphot}, there is no 
statistically significant evidence
for variability from the progenitor.  The root-mean-square
(RMS) of the variability is $1505 L_\odot$ as compared to a
mean square uncertainty of $1204 L_\odot$.  Subtracting
in quadrature, this implies variability at an RMS level of
$900 L_\odot$.  Because the RMS is so close to the level
of the uncertainties, these should be regarded as upper
limits on the variability rather than detections.
If we fit a linear trend, the slope is
$(220\pm 133)L_\odot$~year$^{-1}$ with $\chi^2=16.4$
for $13$ degrees of freedom.  If we rescale the uncertainties
to make the $\chi^2$ per degree of freedom unity,
this becomes $(291\pm 149)L_\odot$~year$^{-1}$.
If we adopt an R-band luminosity of $\nu L_\nu = 10^{4.3}L_\odot$,
the RMS variability is 
$< 4.5\%$ and the limit on the slope is 
$(1.4\pm 0.7)\%$~year$^{-1}$.
This luminosity estimate is the logarithmic average of the Galactic extinction-corrected
ACS and WFC F814W and F555W C1, D1 and E1 source luminosities.
The absolute scale of these luminosities is subject to
distance uncertainties and extinction corrections, 
but the fractional variability is independent of both.

Figure~\ref{fig:vary1} shows the relation of the LBT epochs
to nuclear burning phases with a gray band bracketing the
onset of core oxygen burning for models with 
$14 M_\odot < M < 20 M_\odot$ from \cite{Sukhbold2014}, where 
the full mass dependence is shown in Figure~\ref{fig:bigpicture}.  
More or less independent of mass, the LBT data samples the last 
phases of carbon shell burning, the neon burning phase (not shown),
core oxygen burning, and the initial phases of shell oxygen 
burning.  Since the two points that drive the apparent slope
in the luminosity evolution probably lie in the 
shell carbon burning phase, the variability in the later
burning phases is even smaller than the overall limits given
above.  

\section{Discussion}

We have clearly identified a counterpart to ASASSN-16fq
in archival HST data. Unfortunately,
our constraints on the properties of the progenitor are
far from satisfactory, presumably due to the blending
of multiple stars with the progenitor even at the resolution of
HST. However, for a broad range of reasonable assumptions about its
temperature and the amount of extinction, its properties are
consistent with a lower-mass ($8$-$12M_\odot$) red supergiant.
The data almost certainly require an upper mass limit
of $M_* \ltorder 17 M_\odot$ that matches the mass limit
associated with the red supergiant problem (\citealt{Smartt2009b})
or the more general problem of missing higher mass SN
progenitors originally identified by \cite{Kochanek2008}.
The only real escape from this conclusion is to give the
progenitor a far higher than expected temperature, but this 
option quickly drives the progenitor luminosity to be
extraordinarily high.  While we view this possibility as
unlikely, it would make the progenitor of ASASSN-16fq far
more remarkable than simply making it a garden variety
red supergiant of modest mass.  Observations either with the LBT
or HST once the SN has faded will have no difficulty making
accurate photometric measurements of the progenitor.
 
Table~\ref{tab:progvary} summarizes the available information on
the variability of SN progenitors beyond the large outbursts 
probed by  \cite{Ofek2014}, \cite{Bilinski2015} and \cite{Strotjohann2015}.
Information is available for the Type~IIpec SN~1987A 
(photographic, \citealt{Plotkin2004}),
the Type~IIb SN~1993J (V-band, \citealt{Cohen1995}), the Type~IIP 
SN~2008cn (V-band, \citealt{EliasRosa2009}, \citealt{Maund2015}),
the Type~IIb SN~2011dh (R-band, \citealt{Szczygiel2012}) and the
Type~IIP SN~2013ej (I-band, \citealt{Fraser2014}) in
addition to ASASSN-16fq.  
Based on the review of \cite{Smartt2009},
we adopt progenitor masses of $14$-$20M_\odot$ for SN~1987A
and $15 M_\odot$ for SN~1993J. We use an upper limit of $<16 M_\odot$ for
SN~2008cn following \cite{Maund2015}, $(13\pm 3)M_\odot$
for SN~2011dh following \cite{Maund2011}, and $8$-$15.5M_\odot$
for SN~2013ej (\citealt{Fraser2014}).  We use our
estimate of $8$-$12 M_\odot$ from \S2 for ASASSN-16fq.
We also report the time period spanned by the variability
data and (roughly) the corresponding nuclear burning
phases based on Figure~\ref{fig:bigpicture}.  The
data for SN~1987A, SN~1993J, SN~2008cn and SN~2013ej
probably only sample the carbon shell burning phase.
The LBT data for SN~2011dh, like that for ASASSN-16fq,
probably samples the last phases of carbon shell burning
through the early phases of oxygen shell burning.

We can characterize the ``random,'' ``steady'', and ``outburst''
variability of these SN progenitors. Limits on the random variability
are illustrated by the ``Var'' estimates of the intrinsic variability
as a function of the progenitor mass in Figure~\ref{fig:vary2}.  
The variance ``Var'' is estimated by subtracting the mean of the 
reported photometric uncertainties ($\langle\hbox{Err}\rangle$)
from the root mean square (RMS) of the light curve,
($\hbox{Var}=(\hbox{RMS}^2-\langle\hbox{Err}\rangle^2)^{1/2}$).
The intrinsic variability is defined to be zero if
the mean errors exceed the RMS, as is the case for
SN~2013ej and the \cite{Maund2015} results for SN~2008cn. 
These quantities are reported in Table~\ref{tab:progvary}.
Limits on the steady variability are illustrated in 
Figure~\ref{fig:vary3} by the estimates of the linear
luminosity slopes as a function of progenitor mass.  The upper limits 
used for all but SN~2011dh are drawn at the absolute value 
of the slope plus the error estimate.  
The slope
estimates and the goodness of fit are included in
Table~\ref{tab:progvary}.   

Of these SN, only SN~2011dh is clearly
variable, but with the small number of epochs available
to \cite{Szczygiel2012} it is also possible to interpret
it as ellipsoidal variability given the binary models
for the progenitor system by \cite{Benvenuto2013}.
For comparison, typical slopes estimated from the end points of stellar
evolution models are $10^{-3}$ to $10^{-4}$~mag/year
(e.g., \citealt{Schaller1992}, \citealt{Heger2000}).  
The limit on the slope for SN~1987A is by far
the tightest due to the long time span of the data.  
Obviously, these systems are 
heterogeneously selected and sample different final
burning phases (see Figure~\ref{fig:bigpicture}), but they 
also appear to be the only
published progenitors with adequate data to test for these lower
levels of variability.  

We can characterize outbursts by adding Gaussian
bursts in luminosity (quadratic in magnitude) defined
by a peak luminosity $L_{peak}$ and a burst FWHM $t_{peak}$
to the light curves of all the sources in Table~\ref{tab:progvary}
except SN~1987A.
We allow the outbursts during an eruption time corresponding
to the last $t_{out}$ before the SN. 
We normalize the available light curves to have a
$\chi^2$ per degree of freedom, $N_{dof}$, of at most unity (i.e.
ignoring the variability of SN~2011dh) when fit as having a
constant flux. We then add model
outbursts at random times and conservatively define detection
to be when the $\chi^2$ for fitting the ``fake'' data containing an 
outburst as having a constant flux
exceeds the larger of $2N_{dof}$ and $N_{dof}+4$. 
Figure~\ref{fig:outburst} shows the results, quantified
as the detection probability per SN, $P_d$, for peak
outburst luminosities of $M_R=-6$, $-8$, $-10$ and $-12$~mag,
corresponding to $\lambda L_\lambda \simeq 10^{4.0}$, $10^{4.8}$
$10^{5.6}$ and $10^{6.4}L_\odot$.  The detection probabilities
can then be converted to 90\% confidence limits on the number
of outbursts per SN as 
\begin{equation}
   N_{out} < { 2.30  \over N_{SN} P_d } = 0.46 P_d^{-1}
\end{equation}
where $N_{SN}=5$ since we have excluded SN~1987A from the 
analysis.  The rate of eruptions during the eruption period 
is then $r_{out} = N_{out}/t_{out}$. 

The general pattern of the detection probabilities
$P_d$  in Figure~\ref{fig:outburst}
is relatively easy to understand.  Short outbursts become
increasingly difficult to detect because of the finite 
temporal sampling of the data.   Long outbursts ultimately
become difficult to detect because they show no time
variability over the finite temporal extent 
of the data (although for sufficiently bright transients
the luminosity would be incompatible with any progenitor).  
The results for long outbursts eventually correspond to 
the slope limits of Figure~\ref{fig:vary3}.
The sensitivity is highest for eruption
periods extending to roughly $10$~years because SN~1993J,
SN~2008cn and SN~2013ej only contribute on these time scales.
For the shortest eruption periods ($t_{out}$), only SN~2011dh contributes,
and for long eruption periods, there is no information 
outside the last roughly $10$~years before the SN.     

Figure~\ref{fig:outburst2} shows outburst limits computed in the
same manner for the sample of $27$ Type~IIb SN considered by
\cite{Strotjohann2015} for comparison.  \cite{Ofek2014} and
\cite{Strotjohann2015} provide $5\sigma$ R-band luminosity
limits, $L_{PTF}$, for 15~day bins of the data, each
containing a variable number, $N_{PTF}$, of epochs. For
simplicity, we simply spread the reported number of epochs 
uniformly over their 15 day bin (with temporal spacings of 
$1/2:1:1 \cdots 1:1:1/2$ over the bin), each with a ($1\sigma$) 
uncertainty per epoch of $L_{PTF}N_{PTF}^{1/2}/5$.  We can then
apply our formalism with only minor ambiguities for
very short time scale ($t_{peak} \ltorder 15$~day) outbursts.  Figure~\ref{fig:outburst2}
shows the results for the \cite{Strotjohann2015} sample at $M_R=-12$ and $-10$~mag.
The Type~IIn sample considered by \cite{Ofek2014} has even less sensitivity to low
luminosity outbursts because the typical SN is more distant and there
are fewer SN in the sample.

As we can see from comparing Figures~\ref{fig:outburst} and \ref{fig:outburst2},
the PTF sample is more sensitive to very luminous outbursts and far less
sensitive to outbursts closer to the progenitor luminosity.  This is simply
because, relative to the sample in Table~\ref{tab:progvary}, the PTF data 
has more continuous, but very shallow, coverage of a larger number of SN.
The rate limits of the two samples cross 
near peak luminosities of $M_R=-12$~mag ($\lambda L_\lambda \simeq 10^{6.4}L_\odot$), 
where the relative sensitivity depends on the burst duration.  By
$M_R=-10$~mag ($\lambda L_\lambda \simeq 10^{5.6}L_\odot$), the deeper data 
we use here is more sensitive independent of the outburst duration. The
PTF data has negligible sensitivity to the fainter $M_R=-8$ and $-6$~mag
outbursts. 

In essence, the two approaches are complimentary.  SN surveys like PTF
will better constrain the rates of high luminosity, shorter transients --
they will generally have larger numbers of SN observed with higher
cadence.  However, even with
the co-addition of data, SN surveys simply lack the sensitivity to  
probe variability significantly below $\sim 10$ times the 
luminosity of the progenitor stars.  Deep monitoring data, like that from our LBT survey, are 
sensitive to very low levels of variability (down to $\sim 1\%$ of
the progenitor luminosity, Figure~\ref{fig:vary1}), but are limited by 
the SN rate in nearby galaxies ($\sim 1$~year) and the lower cadence 
of any monitoring project on large telescopes.  

SN surveys like PTF are also largely limited to studying the relationships between outburst
and SN properties, as done by \cite{Ofek2014}, because most of the SN 
will be too distant for measurements of the progenitor properties.  
Any survey which can measure variability on the scale of the progenitor
luminosity or fainter can, by definition, also determine the properties
of the progenitor.  As a result, studies like the LBT survey (\citealt{Kochanek2008},
\citealt{Gerke2015}) are
better suited to studying the relationship between outbursts and 
progenitors.  An obvious next step is to systematically analyze the variability
of all the SN progenitors in the LBT survey data, which will provide
a relatively homogeneous, volume-limited sample, rather than the 
heterogeneous sample represented by Table~\ref{tab:progvary}. 

\section*{Acknowledgments}

We thank A. Dolphin for his advice regarding {\sc dolphot}.
CSK, KZS, JSB, SMA and TWSH are supported by NSF grants AST-1515876 and
AST-1515927.  
BJS is supported by NASA through Hubble Fellowship 
grant HF-51348.001 awarded by the Space Telescope Science Institute, 
which is operated by the Association of Universities for Research in 
Astronomy, Inc., for NASA, under contract NAS 5-26555.   
TW-SH is supported by the DOE Computational
Science Graduate Fellowship, grant number DE-FG02-
97ER25308.  TS is partly supported by NSF grand PHY-1404311 to J. Beacom.
This work was partly supported by the European Union 
FP7 program through ERC grant number 320360. 
Support for JLP is provided in part by FONDECYT through the
grant 1151445 and by the Ministry of Economy, Development, and Tourism's Millennium Science Initiative
through grant IC120009, awarded to The Millennium Institute of
Astrophysics, MAS.
SD is supported by the Strategic Priority Research Program ``The
Emergence of Cosmological Structures'' of the Chinese Academy of
Sciences (Grant No. XDB09000000) and NSFC project 11573003.
Some of the observations were carried out using the LBT at Mt Graham, AZ. The LBT
is an international collaboration among institutions in the United
States, Italy, and Germany. LBT Corporation partners are the University
of Arizona on behalf of the Arizona university system; Istituto
Nazionale di Astrofisica, Italy; LBT Beteiligungsgesellschaft,
Germany, representing the Max-Planck Society, the Astrophysical
Institute Potsdam, and Heidelberg University; the Ohio State
University; and The Research Corporation, on behalf of the University
of Notre Dame, University of Minnesota, and University of
Virginia.
This work is based in part on observations
made with the Spitzer Space Telescope, which is operated
by the Jet Propulsion Laboratory, California Institute of Technology
under a contract with NASA, and in part on observations made
with the NASA/ESA Hubble Space Telescope obtained at the Space
Telescope Institute, which is operated by the Association of Universities
for Research in Astronomy, Inc., under NASA contract
NAS 5-26555. Some observations were obtained from the Hubble Legacy Archive, which is a 
collaboration between the Space Telescope Science Institute (STScI/NASA), 
the Space Telescope European Coordinating Facility (ST-ECF/ESA) and 
the Canadian Astronomy Data Centre (CADC/NRC/CSA).

\vfill\eject

\begin{table*}
   \centering
   \caption{HST photometry}
   \begin{tabular}{llllccccc}
   \hline
Date          &Instr.         &Exp.            &Src.       & F435W               & F555W               & F606W                & F658N                & F814W    \\
              &               &(\#$\times$sec) &           & (mag)               & (mag)               & (mag)                & (mag)                & (mag)    \\
   \hline
1994-12-28    &WFPC2          & 2$\times$80    & All       & --                  & --                  &23.639$\pm$0.099      & --                   &  --      \\
2001-02-24    &WFPC2          & 2$\times$350   & All       & --                  & --                  & --                   & --                   &22.225$\pm$0.041    \\
              &               &                & HSC       & --                  & --                  & --                   & --                   &22.302$\pm$0.083     \\
2001-03-04    &WFPC2          & 2$\times$350   & All       & --                  &23.396$\pm$0.083     & --                   & --                   & -- \\
2001-11-26    &WFPC2          &160$+$400       & All       & --                  & --                  &23.160$\pm$0.038      & --                   & -- \\  
2004-12-31    &ACS            & 2$\times$600   & B1        & --                  & --                  & --                   &22.708$\pm$0.081      & --  \\
              &               &                & B2        & --                  & --                  & --                   &24.356$\pm$0.315      & --  \\ 
              &               &                & B3        & --                  & --                  & --                   &24.384$\pm$0.323      & --  \\
              &               &                & HSC       & --                  & --                  & --                   &22.813$\pm$0.075      & --  \\
2004-12-31    &ACS            & 2$\times$500   & A1        &24.149$\pm$0.041     & --                  & --                   & --                   & --  \\
              &               &                & A2        &26.034$\pm$0.190     & --                  & --                   & --                   & --  \\
              &               &                & A3        &26.011$\pm$0.180     & --                  & --                   & --                   & --  \\
              &               &                & A4        &26.250$\pm$0.210     & --                  & --                   & --                   & --  \\
              &               &                & A5        &26.189$\pm$0.187     & --                  & --                   & --                   & --  \\
              &               &                & A6        &26.727$\pm$0.282     & --                  & --                   & --                   & --  \\
              &               &                & HSC       &24.097$\pm$0.024     & --                  & --                   & --                   & --  \\
2009-12-14    &ACS            & 2$\times$260   & C1        & --                  &23.899$\pm$0.037     & --                   & --                   & --  \\
              &               &                & C2        & --                  &25.015$\pm$0.085     & --                   & --                   & --  \\
              &               &                & C3        & --                  &25.392$\pm$0.114     & --                   & --                   & --  \\
              &               &                & C4        & --                  &26.259$\pm$0.233     & --                   & --                   & --  \\
              &               &                & C5        & --                  &26.889$\pm$0.416     & --                   & --                   & --  \\
              &               &                & C6        & --                  &26.254$\pm$0.233     & --                   & --                   & --  \\
              &               &                & C7        & --                  &26.119$\pm$0.196     & --                   & --                   & --  \\
              &               &                & HSC       & --                  &23.768$\pm$0.032     &--                    & --                   & --  \\
2009-12-14    &ACS            & 2$\times$260   & D1        & --                  & --                  & --                   & --                   &23.308$\pm$0.036 \\
              &               &                & D2        & --                  & --                  & --                   & --                   &23.720$\pm$0.050 \\
              &               &                & D3        & --                  & --                  & --                   & --                   &24.224$\pm$0.073 \\
              &               &                & D4        & --                  & --                  & --                   & --                   &24.991$\pm$0.133 \\
              &               &                & HSC       & --                  &  --                 &  --                  & --                   &22.721$\pm$0.022 \\
2013-11-29    &WFC3           & 3$\times$373   & E1        & --                  &24.158$\pm$0.028     & --                   & --                   & --  \\
              &               &                & E2        & --                  &25.202$\pm$0.064     & --                   & --                   & --  \\
              &               &                & E3        & --                  &25.967$\pm$0.108     & --                   & --                   & --  \\
              &               &                & E4        & --                  &26.515$\pm$0.160     & --                   & --                   & --  \\
2013-11-29    &WFC3           & 3$\times$373   & E1        &  --                 & --                  & --                   & --                   &23.204$\pm$0.033  \\
              &               &                & E2        & --                  & --                  & --                   & --                   &23.919$\pm$0.060  \\
              &               &                & E3        & --                  & --                  & --                   & --                   &25.044$\pm$0.139  \\
              &               &                & E4        & --                  & --                  & --                   & --                   &24.140$\pm$0.064  \\
   \hline 
   \multicolumn{9}{l}{The detectors for the WFPC2, ACS and WFC3 instruments were WF3, WFC1 and UVIS2, respectively.  The WFPC2/WF3 } \\
   \multicolumn{9}{l}{photometry encompasses ``All'' of the sources.  For WFPC2/WF3 (ACS/WFC1) the HSC AB TotMag (MagAp2) results } \\
   \multicolumn{9}{l}{were converted to Vega magnitudes. }
   \end{tabular}
   \label{tab:hstphot}
\end{table*}

\vfill\eject

\def\sp{\hphantom{0}}
\begin{table*}
   \centering
   \caption{LBT R-Band Variability}
   \begin{tabular}{llr}
     \multicolumn{1}{c}{Date}
    &\multicolumn{1}{c}{MJD}
    &\multicolumn{1}{c}{$\Delta L_R$ ($L_\odot$)} \\
   \hline
      2008-05-04 &  54590.24 &$184 \pm  1236$ \\ 
2009-01-30 &  54861.34 &$-2977 \pm  1553$ \\ 
2009-03-22 &  54912.25 &$-2335 \pm  1641$ \\ 
2010-12-13 &  55543.44 &$-1771 \pm  1570$ \\ 
2012-01-01 &  55927.44 &$-1068 \pm  1010$ \\ 
2012-03-22 &  56008.26 &$-221 \pm \sp 987$ \\ 
2012-04-28 &  56045.18 &$3041 \pm  1132$ \\ 
2013-03-16 &  56367.29 &$-228 \pm  1176$ \\ 
2013-05-05 &  56417.21 &$330 \pm \sp 872$ \\ 
2014-01-09 &  56666.36 &$549 \pm  1273$ \\ 
2014-04-25 &  56772.20 &$-371 \pm \sp 734$ \\ 
2015-01-19 &  57041.32 &$-571 \pm  1323$ \\ 
2015-04-20 &  57132.23 &$1604 \pm \sp 975$ \\ 
2016-01-03 &  57390.44 &$-80 \pm  1064$ \\ 
2016-02-07 &  57425.29 &$496 \pm  1116$ \\ 

   \hline
   \end{tabular}
   \label{tab:lbtphot}
\end{table*}

\def\sp{\hphantom{0}}
\begin{table*}
   \centering
   \caption{Progenitor Variability}
   \begin{tabular}{lccccccrl}
     \multicolumn{1}{c}{SN}
    &\multicolumn{1}{c}{Time}
    &\multicolumn{1}{c}{Burning}
    &\multicolumn{1}{c}{RMS}
    &\multicolumn{1}{c}{$\langle\hbox{Err}\rangle$}
    &\multicolumn{1}{c}{Var}
    &\multicolumn{1}{c}{Slope}
    &\multicolumn{1}{c}{$\chi^2/\hbox{dof}$}
    &\multicolumn{1}{c}{Reference}\\

    &\multicolumn{1}{c}{(years)}
    &\multicolumn{1}{c}{Phase}
    &\multicolumn{1}{c}{(mag)}
    &\multicolumn{1}{c}{(mag)}
    &\multicolumn{1}{c}{(mag)}
    &\multicolumn{1}{c}{(mag/year)}
    &
    & \\
   \hline
      SN~1987a    &91-33     &C Shell        &0.3\sp\sp &?        &?          &\sp$0.005 \pm ??$\sp    &--           &\cite{Plotkin2004} \\
SN~1993J    &9.2-8.8   &C Shell       &0.17\sp   &0.16\sp  &0.05\sp    &$-0.08\sp \pm 0.25$\sp  &53.3/49      &\cite{Cohen1995} \\
SN~2008cn   &12.0-10.8 &C Shell       &0.22\sp   &0.13\sp  &0.18\sp    &$-0.29\sp \pm 0.10$\sp  &14.2/{\sp}7  &\cite{EliasRosa2009} \\
            &          &              &0.13\sp   &0.16\sp  &0\sp\sp\sp &$-0.03\sp \pm 0.12$\sp  &6.1/{\sp}7   &\cite{Maund2015} \\
SN~2011dh   &3.2-0.1   &C$\rightarrow$O Shell        &0.046     &0.022    &0.040      &\sp$0.039 \pm 0.006$    &6.7/{\sp}9   &\cite{Szczygiel2012} \\
SN~2013ej   &9.7-8.1   &C shell        &0.0\sp\sp &0.03\sp  &0\sp\sp\sp &\sp$0.00  \pm 0.026$    &0.0/{\sp}0   &\cite{Fraser2014} \\ 
ASASSN-16fq &8.1-0.3   &C$\rightarrow$O Shell &0.081     &0.065    &0.049      &$-0.015   \pm 0.008$    &16.4/13      &This paper \\

   \hline
   \end{tabular}
   \label{tab:progvary}
\end{table*}


\begin{thebibliography}{}
\bibitem[Abbott et al.(2016)]{Abbott2016} Abbott, B.~P., Abbott, R., Abbott, T.~D., et al.\ 2016, Physical Review Letters, 116, 061102 
\bibitem[Adams \& Kochanek(2015)]{Adams2015} Adams, S.~M., \& Kochanek, C.~S.\ 2015, \mnras, 452, 2195 
\bibitem[Adams et al.(2016)]{Adams2016} Adams, S.~M., Kochanek, C.~S., Prieto, J.~L., et al.\ 2016, \mnras, 460, 1645 
\bibitem[Ahn et al.(2012)]{Ahn2012} Ahn, C.~P., Alexandroff, R., Allende Prieto, C., et al.\ 2012, \apjs, 203, 21 
\bibitem[Alard \& Lupton(1998)]{Alard1998} Alard, C., \& Lupton, R.~H.\ 1998, \apj, 503, 325 
\bibitem[Arnett et al.(1989)]{Arnett1989} Arnett, W.~D., Bahcall, J.~N., Kirshner, R.~P., \& Woosley, S.~E.\ 1989, \araa, 27, 629 
\bibitem[Belczynski et al.(2016)]{Belczynski2016} Belczynski, K., Holz, D.~E., Bulik, T., \& O'Shaughnessy, R.\ 2016, arXiv:1602.04531 
\bibitem[Benvenuto et al.(2013)]{Benvenuto2013} Benvenuto, O.~G., Bersten, M.~C., \& Nomoto, K.\ 2013, \apj, 762, 74 
\bibitem[Bilinski et al.(2015)]{Bilinski2015} Bilinski, C., Smith, N., Li, W., et al.\ 2015, \mnras, 450, 246 
\bibitem[Bock et al.(2016)]{Bock2016} Bock, G., Dong, S, Kochanek, C.S., et al.\ 2016, The Astronomer's Telegram, 9091  
\bibitem[Bressan et al.(2012)]{Bressan2012} Bressan, A., Marigo, P., Girardi, L., et al.\ 2012, \mnras, 427, 127 
\bibitem[Brown \& Woosley(2013)]{Brown2013} Brown, J.~M., \& Woosley, S.~E.\ 2013, \apj, 769, 99 
\bibitem[Cao et al.(2013)]{Cao2013} Cao, Y., Kasliwal, M.~M., Arcavi, I., et al.\ 2013, \apjl, 775, L7 
\bibitem[Castelli \& Kurucz(2004)]{Castelli2004} Castelli, F., \& Kurucz, R.~L.\ 2004, arXiv:astro-ph/0405087 
\bibitem[Couch \& Ott(2015)]{Couch2015} Couch, S.~M., \& Ott, C.~D.\ 2015, \apj, 799, 5 
\bibitem[Clausen et al.(2015)]{Clausen2015} Clausen, D., Piro, A.~L., \& Ott, C.~D.\ 2015, \apj, 799, 190 
\bibitem[Cohen et al.(1995)]{Cohen1995} Cohen, J.~G., Darling, J., \& Porter, A.\ 1995, \aj, 110, 308 
\bibitem[Dalcanton et al.(2009)]{Dalcanton09} Dalcanton, J.~J., Williams, B.~F., Seth, A.~C., et al.\ 2009, ApJS, 183, 67
\bibitem[Dolence et al.(2015)]{Dolence2015} Dolence, J.~C., Burrows, A., \& Zhang, W.\ 2015, \apj, 800, 10 
\bibitem[Dolphin(2000)]{Dolphin2000} Dolphin, A.~E.\ 2000, \pasp, 112, 1383
\bibitem[Eldridge et al.(2013)]{Eldridge2013} Eldridge, J.~J., Fraser, M., Smartt, S.~J., Maund, J.~R., \& Crockett, R.~M.\ 2013, \mnras, 436, 774 
\bibitem[Elias-Rosa et al.(2009)]{EliasRosa2009} Elias-Rosa, N., Van Dyk, S.~D., Li, W., et al.\ 2009, \apj, 706, 1174 
\bibitem[Elias-Rosa et al.(2011)]{EliasRosa2011} Elias-Rosa, N., Van Dyk, S.~D., Li, W., et al.\ 2011, \apj, 742, 6 
\bibitem[Elitzur \& Ivezi{\'c}(2001)]{Elitzur2001} Elitzur, M., \& Ivezi{\'c}, {\v Z}.\ 2001, \mnras, 327, 403 
\bibitem[Ertl et al.(2016)]{Ertl2016} Ertl, T., Janka, H.-T., Woosley, S.~E., Sukhbold, T., \& Ugliano, M.\ 2016, \apj, 818, 124 
\bibitem[Filippenko(1997)]{Filippenko1997} Filippenko, A.~V.\ 1997, \araa, 35, 309 
\bibitem[Folatelli et al.(2016)]{Folatelli2016} Folatelli, G., Van Dyk, S.~D., Kuncarayakti, H., et al.\ 2016, arXiv:1604.06821 
\bibitem[Foley et al.(2011)]{Foley2011} Foley, R.~J., Berger, E., Fox, O., et al.\ 2011, \apj, 732, 32 
\bibitem[Fraser et al.(2012)]{Fraser2012} Fraser, M., Maund, J.~R., Smartt, S.~J., et al.\ 2012, \apjl, 759, L13 
\bibitem[Fraser et al.(2013)]{Fraser2013} Fraser, M., Magee, M., Kotak, R., et al.\ 2013, \apjl, 779, L8 
\bibitem[Fraser et al.(2014)]{Fraser2014} Fraser, M., Maund, J.~R., Smartt, S.~J., et al.\ 2014, \mnras, 439, L56 
\bibitem[Gal-Yam(2012)]{Galyam2012} Gal-Yam, A.\ 2012, Science, 337, 927 
\bibitem[Gerke et al.(2015)]{Gerke2015} Gerke, J.~R., Kochanek, C.~S., \& Stanek, K.~Z.\ 2015, \mnras, 450, 3289 
\bibitem[Giallongo et al.(2008)]{Giallongo2008} Giallongo, E., Ragazzoni, R., Grazian, A., et al.\ 2008, \aap, 482, 349 
\bibitem[Groh et al.(2013)]{Groh2013} Groh, J.~H., Meynet, G., Georgy, C., \& Ekstr{\"o}m, S.\ 2013, \aap, 558, A131 
\bibitem[Heger \& Langer(2000)]{Heger2000} Heger, A., \& Langer, N.\ 2000, \apj, 544, 1016 
\bibitem[Ivezi{\'c} \& Elitzur(1997)]{Ivezic1997} Ivezi{\'c}, Z., \& Elitzur, M.\ 1997, \mnras, 287, 799 
\bibitem[Ivezi{\'c} et al.(1999)]{Ivezic1999} Ivezi{\'c}, Z., Nenkova, M., \& Elitzur, M.\ 1999, arXiv:astro-ph/9910475 
\bibitem[Jennings et al.(2014)]{Jennings2014} Jennings, Z.~G., Williams, B.~F., Murphy, J.~W., et al.\ 2014, \apj, 795, 170 
\bibitem[Horiuchi et al.(2011)]{Horiuchi2011} Horiuchi, S., Beacom, J.~F., Kochanek, C.~S., et al.\ 2011, \apj, 738, 154 
\bibitem[Humphreys \& Davidson(1994)]{Humphreys1994} Humphreys, R.~M., \& Davidson, K.\ 1994, \pasp, 106, 1025 
\bibitem[Ivezi{\'c} et al.(2007)]{Ivezic2007} Ivezi{\'c}, {\v Z}., Smith, J.~A., Miknaitis, G., et al.\ 2007, \aj, 134, 973 
\bibitem[Jerkstrand et al.(2014)]{Jerkstrand2014} Jerkstrand, A., Smartt, S.~J., Fraser, M., et al.\ 2014, \mnras, 439, 3694 
\bibitem[Jordi et al.(2006)]{Jordi2006} Jordi, K., Grebel, E.~K., \& Ammon, K.\ 2006, \aap, 460, 339 
\bibitem[Kanbur et al.(2003)]{Kanbur2003} Kanbur, S.~M., Ngeow, C., Nikolaev, S., Tanvir, N.~R., \& Hendry, M.~A.\ 2003, \aap, 411, 361 
\bibitem[Khan et al.(2010)]{Khan2010} Khan, R., Stanek, K.~Z., Prieto, J.~L., et al.\ 2010, \apj, 715, 1094 
\bibitem[Khan et al.(2015a)]{Khan2015a} Khan, R., Kochanek, C.~S., Stanek, K.~Z., \& Gerke, J.\ 2015, \apj, 799, 187 
\bibitem[Khan et al.(2015b)]{Khan2015b} Khan, R., Adams, S.~M., Stanek, K.~Z., Kochanek, C.~S., \& Sonneborn, G.\ 2015, \apjl, 815, L18 
\bibitem[Kiewe et al.(2012)]{Kiewe2012} Kiewe, M., Gal-Yam, A., Arcavi, I., et al.\ 2012, \apj, 744, 10 
\bibitem[Kochanek et al.(2008)]{Kochanek2008} Kochanek, C.~S., Beacom, J.~F., Kistler, M.~D., et al.\ 2008, \apj, 684, 1336
\bibitem[Kochanek(2011)]{Kochanek2011} Kochanek, C.~S.\ 2011, \apj, 743, 73 
\bibitem[Kochanek et al.(2011)]{Kochanek2011b} Kochanek, C.~S., Szczygiel, D.~M., \& Stanek, K.~Z.\ 2011, \apj, 737, 76 
\bibitem[Kochanek et al.(2012a)]{Kochanek2012} Kochanek, C.~S., Khan, R., \& Dai, X.\ 2012, \apj, 759, 20 
\bibitem[Kochanek et al.(2012b)]{Kochanek2012b} Kochanek, C.~S., Szczygie{\l}, D.~M., \& Stanek, K.~Z.\ 2012, \apj, 758, 142 
\bibitem[Kochanek(2014)]{Kochanek2014} Kochanek, C.~S.\ 2014, \apj, 785, 28 
\bibitem[Kochanek(2015)]{Kochanek2015} Kochanek, C.~S.\ 2015, \mnras, 446, 1213 
\bibitem[Li et al.(2006)]{Li2006} Li, W., Van Dyk, S.~D., Filippenko, A.~V., et al.\ 2006, \apj, 641, 1060 
\bibitem[Li et al.(2011)]{Li2011} Li, W., Leaman, J., Chornock, R., et al.\ 2011, \mnras, 412, 1441 
\bibitem[Lien et al.(2010)]{Lien2010} Lien, A., Fields, B.~D., \& Beacom, J.~F.\ 2010, \prd, 81, 083001 
\bibitem[Margutti et al.(2014)]{Margutti2014} Margutti, R., Milisavljevic, D., Soderberg, A.~M., et al.\ 2014, \apj, 780, 21 
\bibitem[Mauerhan et al.(2013)]{Mauerhan2013} Mauerhan, J.~C., Smith, N., Filippenko, A.~V., et al.\ 2013, \mnras, 430, 1801 
\bibitem[Maund et al.(2004)]{Maund2004} Maund, J.~R., Smartt, S.~J., Kudritzki, R.~P., Podsiadlowski, P., \& Gilmore, G.~F.\ 2004, \nat, 427, 129 
\bibitem[Maund et al.(2011)]{Maund2011} Maund, J.~R., Fraser, M., Ergon, M., et al.\ 2011, \apjl, 739, L37 
\bibitem[Maund et al.(2015)]{Maund2015} Maund, J.~R., Fraser, M., Reilly, E., Ergon, M., \& Mattila, S.\ 2015, \mnras, 447, 3207 
\bibitem[M{\"u}ller(2016)]{Muller2016} M{\"u}ller, B.\ 2016, arXiv:1608.03274 
\bibitem[O'Connor \& Ott(2013)]{OConnor2013} O'Connor, E., \& Ott, C.~D.\ 2013, \apj, 762, 126 
\bibitem[Ofek et al.(2014)]{Ofek2014}Ofek, E.~O., Sullivan, M., Shaviv, N.~J., et al.\ 2014, \apj, 789, 104 
\bibitem[Ofek et al.(2016)]{Ofek2016} Ofek, E.~O., Cenko, S.~B., Shaviv, N.~J., et al.\ 2016, arXiv:1605.02450 
\bibitem[Pastorello et al.(2007)]{Pastorello2007} Pastorello, A., Smartt, S.~J., Mattila, S., et al.\ 2007, \nat, 447, 829 
\bibitem[Pastorello et al.(2013)]{Pastorello2013} Pastorello, A., Cappellaro, E., Inserra, C., et al.\ 2013, \apj, 767, 1 
\bibitem[Pejcha \& Thompson(2015)]{Pejcha2015} Pejcha, O., \& Thompson, T.~A.\ 2015, \apj, 801, 90 
\bibitem[Pejcha \& Prieto(2015)]{Pejcha2015b} Pejcha, O., \& Prieto, J.~L.\ 2015, \apj, 799, 215 
\bibitem[Pier et al.(2003)]{Pier2003} Pier, J.~R., Munn, J.~A., Hindsley, R.~B., et al.\ 2003, \aj, 125, 1559 
\bibitem[Plotkin \& Clayton(2004)]{Plotkin2004} Plotkin, R.~M., \& Clayton, G.~C.\ 2004, Journal of the American Association of Variable Star Observers (JAAVSO), 32, 89 
\bibitem[Quataert \& Shiode(2012)]{Quataert2012} Quataert, E., \& Shiode, J.\ 2012, \mnras, 423, L92 
\bibitem[Reynolds et al.(2015)]{Reynolds2015} Reynolds, T.~M., Fraser, M., \& Gilmore, G.\ 2015, \mnras, 453, 2885 
\bibitem[Schaller et al.(1992)]{Schaller1992} Schaller, G., Schaerer, D., Meynet, G., \& Maeder, A.\ 1992, A\&AS, 96, 269 
\bibitem[Schlafly \& Finkbeiner(2011)]{Schlafly2011} Schlafly, E.~F., \& Finkbeiner, D.~P.\ 2011, \apj, 737, 103 
\bibitem[Shappee et al.(2014)]{Shappee2014} Shappee, B.~J., Prieto, J.~L., Grupe, D., et al.\ 2014, \apj, 788, 48 
\bibitem[Shiode \& Quataert(2014)]{Shiode2014} Shiode, J.~H., \& Quataert, E.\ 2014, \apj, 780, 96 
\bibitem[Smartt et al.(2004)]{Smartt2004} Smartt, S.~J., Maund, J.~R., Hendry, M.~A., et al.\ 2004, Science, 303, 499 
\bibitem[Smartt(2009)]{Smartt2009} Smartt, S.~J.\ 2009, \araa, 47, 63 
\bibitem[Smartt(2015)]{Smartt2015} Smartt, S.~J.\ 2015, \pasa, 32, e016 
\bibitem[Smartt et al.(2009)]{Smartt2009b} Smartt, S.~J., Eldridge, J.~J., Crockett, R.~M., \& Maund, J.~R.\ 2009, \mnras, 395, 1409 
\bibitem[Smith \& McCray(2007)]{Smith2007} Smith, N., \& McCray, R.\ 2007, \apjl, 671, L17 
\bibitem[Smith et al.(2010)]{Smith2010} Smith, N., Miller, A., Li, W., et al.\ 2010, \aj, 139, 1451 
\bibitem[Smith et al.(2011)]{Smith2011} Smith, N., Li, W., Silverman, J.~M., Ganeshalingam, M., \& Filippenko, A.~V.\ 2011, \mnras, 415, 773 
\bibitem[Smith(2014)]{Smith2014b} Smith, N.\ 2014, \araa, 52, 487 
\bibitem[Smith \& Arnett(2014)]{Smith2014} Smith, N., \& Arnett, W.~D.\ 2014, \apj, 785, 82 
\bibitem[Stancliffe \& Eldridge(2009)]{Stancliffe2009} Stancliffe, R.~J., \& Eldridge, J.~J.\ 2009, \mnras, 396, 1699 
\bibitem[Strotjohann et al.(2015)]{Strotjohann2015} Strotjohann, N.~L., Ofek, E.~O., Gal-Yam, A., et al.\ 2015, \apj, 811, 117 
\bibitem[Sukhbold \& Woosley(2014)]{Sukhbold2014} Sukhbold, T., \& Woosley, S.~E.\ 2014, \apj, 783, 10 
\bibitem[Sukhbold et al.(2016)]{Sukhbold2016} Sukhbold, T., Ertl, T., Woosley, S.~E., Brown, J.~M., \& Janka, H.-T.\ 2016, \apj, 821, 38 
\bibitem[Szczygie{\l} et al.(2012)]{Szczygiel2012} Szczygie{\l}, D.~M., Gerke, J.~R., Kochanek, C.~S., \& Stanek, K.~Z.\ 2012, \apj, 747, 23 
\bibitem[Tinyanont et al.(2016)]{Tinyanont2016} Tinyanont, S., Kasliwal, M.~M., Fox, O.~D., et al.\ 2016, arXiv:1601.03440 
\bibitem[Ugliano et al.(2012)]{Ugliano2012} Ugliano, M., Janka, H.-T., Marek, A., \& Arcones, A.\ 2012, \apj, 757, 69 
\bibitem[Van Dyk et al.(2000)]{VanDyk2000} Van Dyk, S.~D., Peng, C.~Y., King, J.~Y., et al.\ 2000, \pasp, 112, 1532 
\bibitem[Van Dyk et al.(2003)]{VanDyk2003} Van Dyk, S.~D., Li, W., \& Filippenko, A.~V.\ 2003, \pasp, 115, 1289 
\bibitem[Van Dyk et al.(2011)]{VanDyk2011} Van Dyk, S.~D., Li, W., Cenko, S.~B., et al.\ 2011, \apjl, 741, L28 
\bibitem[Walmswell \& Eldridge(2012)]{Walmswell2012} Walmswell, J.~J., \& Eldridge, J.~J.\ 2012, \mnras, 419, 2054 
\bibitem[Whitmore et al.(2016)]{Whitmore2016} Whitmore, B.~C., Allam, S.~S., Budav{\'a}ri, T., et al.\ 2016, \aj, 151, 134 
\bibitem[Woosley et al.(2002)]{Woosley2002} Woosley, S.~E., Heger, A., \& Weaver, T.~A.\ 2002, Reviews of Modern Physics, 74, 1015 
\bibitem[Woosley \& Heger(2007)]{Woosley2007} Woosley, S.~E., \& Heger, A.\ 2007, \physrep, 442, 269 
\bibitem[Wongwathanarat et al.(2015)]{Wongwathanarat2015} Wongwathanarat, A., M{\"u}ller, E., \& Janka, H.-T.\ 2015, \aap, 577, A48 
\bibitem[Woosley \& Heger(2015)]{Woosley2015} Woosley, S.~E., \& Heger, A.\ 2015, \apj, 810, 34 
\bibitem[Woosley(2016)]{Woosley2016} Woosley, S.~E.\ 2016, arXiv:1603.00511 
\bibitem[Zhang et al.(2016)]{Zhang2016} Zhang, J., Zheng, X, Wang, X, \& Rui, L., 2016, The Astronomer's Telegram, 9093  
\end{thebibliography}
\end{document}